\documentclass[useAMS,usenatbib]{mnras}

\usepackage{graphicx}
\usepackage{color}
\usepackage{multirow}
\usepackage{amssymb,amsmath}
\usepackage{cprotect}

\def\aap{A\&A}
\def\apj{ApJ}
\def\araa{ARAA}
\def\apjl{ApJL}

\def\pasp{PASP}
\def\pasa{PASA}
\def\mnras{MNRAS}

\def\apjs{ApJS}

\def\ps{$\rm km \,s^{-1}\,kpc^{-1}$}

\def\kms{\,km\,s$^{-1}$}

\def\H2{$\rm H_2$}

\title[GMCs in tidal spirals]{The changing GMC population in galaxy interactions}

\author[A. R. Pettitt et al.]
{Alex R. Pettitt$^{1}$\thanks{E-mail:
alex@astro1.sci.hokudai.ac.jp}, Fumi Egusa$^{2}$, Clare L. Dobbs$^3$, Elizabeth J. Tasker$^4$, \and Yusuke Fujimoto$^5$ and Asao Habe$^{1}$\\
$^{1}$Department of Physics, Faculty of Science, Hokkaido University, Sapporo 060-0810, Japan\\
$^{2}$Chile Observatory, National Astronomical Observatory of Japan, 2-21-1 Osawa, Mitaka, Tokyo 181-8588, Japan\\
$^{3}$School of Physics and Astronomy, University of Exeter, Stocker Road, Exeter, EX4 4QL, UK\\
$^{4}$Institute of Space and Astronomical Science, Japan Aerospace Exploration Agency, Yoshinodai 3-1-1, Sagamihara, Kanagawa, Japan\\
$^{5}$The Australian National University, Canberra, ACT 0200, Australia\\
}

\begin{document}
\date{\today}

\pagerange{\pageref{firstpage}--\pageref{lastpage}} \pubyear{201X}

\maketitle

\label{firstpage}

\begin{abstract}
{
With the advent of modern observational efforts providing extensive giant molecular cloud catalogues, understanding the evolution of such clouds in a galactic context is of prime importance. While numerous previous numerical and theoretical works have focused on the cloud properties in isolated discs, few have looked into the cloud population in an interacting disc system. We present results of the first study investigating the evolution of the cloud population in galaxy experiencing an M51-like tidal fly-by using numerical simulations including star formation, interstellar medium cooling and stellar feedback. We see the cloud population shift to large unbound clouds in the wake of the companion passage, with the largest clouds appearing as fleeting short-lived agglomerations of smaller clouds within the tidal spiral arms, brought together by large scale streaming motions. These are then sheared apart as they leave the protection of the spiral arms. Clouds appear to lead diverse lives, even within similar environments, with some being born from gas shocked by filaments streaming into the spiral arms, and others from effectively isolated smaller colliding pairs. Overall this cloud population produces a shallower mass function than the disc in isolation, especially in the arms compared to the inter-arm regions. Direct comparisons to M51 observations show similarities between cloud populations, though models tailored to the mass and orbital models of M51 appear necessary to precisely reproduce the cloud population.
}
\end{abstract}


\section{Introduction}
Star formation within galaxies is believed to take place predominantly within giant molecular clouds (GMCs). These massive complexes of dense gas are observed throughout the Milky Way and external galaxies (e.g. \citet{2007ApJ...654..240R}, \citealt{2014ApJ...784....3C}, \citealt{2016ApJ...822...52R}). As they act as the primary star formation engines of galaxies, it is of paramount importance to understand their formation and evolution, including the possible influence of the wider galactic-scale environment.

Observations of gas in our Milky Way galaxy offers the highest resolution GMC sample, though are hindered by our incomplete picture of the global spiral and bar structure. There have been numerous cloud catalogues of the Milky Way over the years over the visible galactic disc \citep{1986ApJ...305..892D,1989ApJ...339..919S, 2001ApJ...551..852H,2009ApJ...699.1092H}, see also the review of \citet{2015ARA&A..53..583H}. GMCs are generally seen to trace out spiral structure, particularly in the nearby Carina and Sagittarius arms, with the distribution becoming more complex close to the inner bar. Other arms such as the Perseus arm however, are traced much poorer by GMCs in certain regions \citep{2010ApJ...723..492R}. Recently \citet{2016ApJ...822...52R} composed a catalogue of over 1000 GMCs that indicated the population changes with galactic radius, even seeing properties vary between galactic quadrants (e.g. cloud mass spectra slopes), likely a result of the complex barred spiral nature of the Milky Way. \citet{2017ApJ...834...57M} create a huge cloud catalogue numbering in over 8000. They find that clouds are loosely associated with spiral arms and they also tend to be less well bound in both the inner disc (where the bar dominates) and, curiously, in the 3$^{\rm rd}$ quadrant.

Clouds in the LMC are seen to be associated with star forming complexes, indicating continued star formation, and show similar scaling relations as those in the Milky Way \citep{1999PASJ...51..745F,2008ApJS..178...56F}. More recently \citet{2010MNRAS.406.2065H} observed a weak dependence of cloud velocity dispersion with gas surface density (see also \citealt{2011ApJS..197...16W}). A nearby disc galaxy neighbour, M33, with a nearly face-on projection, has also been well observed in many studies \citep{2003ApJ...599..258R,2003ApJS..149..343E,2007ApJ...661..830R,2012A&A...542A.108G,2017A&A...601A.146C,2018A&A...612A..51B}. No large asymmetry between arm and inter-arm cloud populations is seen in M33, though some north-south and inner-outer differences are seen \citep{2007ApJ...661..830R,2012A&A...542A.108G}. The borderline flocculent nature of M33 makes it difficult to trace changes in the GMC population back to large-scale galactic dynamics \citep{2018MNRAS.478.3793D}.

Observations have also been conducted of a range of galactic types, such as spiral armed (M31, \citealt{2007ApJ...654..240R}), dusty (M64, \citealt{2005ApJ...623..826R}) and lenticular galaxies (NGC\,4526, \citealt{2015ApJ...803...16U}). While M31 appears to have similar cloud populations to the Milky Way, NGC\,4526 displays a population that does not trace the scaling relations seen in the Galaxy \citep{1981MNRAS.194..809L}. The emergence of modern facilities such as the Atacama Large Millimeter Array has brought with it studies of GMCs across a much greater variety of galaxies. Observations of barred spiral galaxies shows clouds in the barred/nuclear regions have higher surface densities compared to their arm/inter-arm brethren (\citealt{2017MNRAS.468.1769F,2017ApJ...839..133P,2018ApJ...854...90E,Hirota18acpt}), with the galaxy as a whole containing a diverse cloud population (see also \citealt{2017PASJ...69...18T,2017ApJ...839....6W}).

The M51 galaxy in particular has been a key observational target; being nearby, face-on, and the ``poster-child" of grand design spiral structure. \citet{2009ApJ...700L.132K} and \citet{2012ApJ...761...41K} show a clear evolutionary picture of clouds moving from inter-arm to arm regions, with streaming motions bringing clouds together into large associations that are sheared apart as they move downstream of the arms \citep{2011ApJ...726...85E}. The PAWS survey of M51 \citep{2013ApJ...779...42S} has presented some of the highest resolution observations of extragalactic GMCs in \citet{2014ApJ...784....3C}. They identify over 1500 clouds and thoroughly investigate the changes in GMCs in a number of distinct regions. They see an environmental dependance between spiral and inter-arm structures, especially in slopes of the GMC mass function, favouring larger clouds in the arm regions. Clouds in the arms are also seen to have slightly higher velocity dispersions compared to inter-arm clouds, with a modest population of unbound clouds. \citet{2017ApJ...836..175K} reproduce the analytic mass functions seen in the PAWS data by assigning different formation timescales in arms and inter-arms as a result of feedback from HII regions, though do not take into account the effect of large scale streaming and shearing flow.

Outside of M51, few interacting galaxies have been subject to a GMC analysis. \citet{2012ApJ...750..136W} detect a number of GMCs in the overlap region of the Antennae galaxies (see also \citealt{2012ApJ...745...65U}), displaying high velocity dispersions resulting from the extreme environment of such a collision (e.g. increased external pressure and frequency of cloud cloud collisions). \citet{2017ApJ...841...43E} study the interacting galaxy pair IC 2163 and NGC 2207, and see a changing GMC population both across each galaxy and between the two.

While observational efforts push closer into the pc regime, so do modern numerical simulations of whole galaxies, allowing for modelling of both cloud scales and the larger galactic scale dynamics. Both grid and particle-based simulations of entire galactic discs have been employed in past studies, all emphasising the need for stellar feedback to regulate cloud growth \citep{2011MNRAS.417.1318D,2015ApJ...801...33T,2016MNRAS.461.1684F,2018MNRAS.479.3167G}. Cloud-cloud collisions have been given particular attention by a number of works \citep{2009ApJ...700..358T,2014MNRAS.445L..65F,2015MNRAS.446.3608D} and their influence on the global cloud population.

As for the structure of a given galaxy, simulations tend to show a definite influence on GMC properties. \citet{2012MNRAS.421.3488H} and \citet{2018MNRAS.475...27N} conduct high resolution simulations of a number of different mass model galaxies (using live stars and static potentials) and in general do not find huge differences between the GMC properties, which may be naively expected from the changes in shear in different rotation curves. Spiral arms and bars, however, appear to play a more prominent role. Simulations of steadily rotating potentials representing bars and arms indicate that they are crucial in gathering gas into larger GMCs \citep{2011MNRAS.417.1318D,2014MNRAS.445L..65F}. Simulations also tend to produce a significant population of unbound clouds \citep{2012MNRAS.421.3488H,2016MNRAS.455.1782K}, though \citet{2011MNRAS.413.2935D} suggest that only small regions within clouds need be bound for the cloud to form stars, the rest existing as an unbound envelope. Recently \citet{2017MNRAS.464..246B} performed an analysis of GMCs in a galaxies with a steady density wave like spiral perturbation, and a dynamic spiral arm model. The population of GMCs is remarkably similar between the two spiral generation mechanisms, despite their very different origins. The main difference appears to be how the GMCs are destroyed; as clouds leave density wave spiral arms they are sheared out into spur features perpendicular to the arms, which is not seen in dynamic arms due to their co-rotation with the disc material.

It is apparent therefore that the spiral and bar structure of galaxies have the capacity to directly influence the properties and evolution of GMCs within. However, there is a deficit in studies looking into a specific spiral generation mechanism: that of tidally induced 2-armed spirals. While at first these may be a niche category of galaxy, it is of utmost importance in terms of our modern understanding of spiral galaxies as it offers the primary mechanism for generating unbarred 2-armed spirals, which are comparatively common in nature (see the review of spiral arms of \citealt{2014PASA...31...35D}). The M51 galaxy, one of the subjects for some of the highest resolution extragalactic GMC populations, is one such galactic system. Numerous simulation efforts have looked into interacting galaxies and their impact on the structure of stars and the ISM \citep{1972ApJ...178..623T,2008ApJ...683...94O,2010MNRAS.403..625D,2013MNRAS.430.1901H,2017MNRAS.468.4189P}. However, none have looked in detail at the GMC properties in such systems, especially with the resolution and physics required to make inferences on the GMC population. In this study we aim to rectify this shortcoming by analysing the evolution and properties of GMCs in a simulation of a galactic flyby tidal encounter.

This paper is organised as follows. In Section\;\ref{SecMeth} we summarise the interacting simulation and our GMC extraction and analysis pipeline. In Section\;\ref{SecRes} we discuss the global results, including how clouds change as a function of time and location in the disc after companion passage. Section\;\ref{SecDisc} includes a comparison of our simulation results to M51 data. We then conclude in Section\;\ref{SecConc}. We remind the reader that this simulation is not meant to be a perfect match to M51, and this work focuses on how the GMC population changes when a galaxy is tidally perturbed. The orbital configuration, mass/velocity model for the galaxy are not tailored to exactly reproduce the M51 system, and M51 data is used as only a qualitative comparison check.

\section{Methodology}
\label{SecMeth}

\subsection{Numerical simulation}
\label{sec:numerics}
We utilise the same simulation as in \citet{2017MNRAS.468.4189P} (hereafter P2017), where we performed calculations of a galactic disc under the influence of the passage of a small companion using the \textsc{gasoline2} code \citep{2017MNRAS.471.2357W}. A bar-stable galaxy is initialised with live stellar disc and bulge, dark matter halo and gas disc, the latter is subject to ISM heating and cooling, star formation and stellar feedback physics. After 400\,Myr of isolated evolution a companion galaxy (a spheroidal distribution of dark matter) approaches perigalacticon position, driving a 2-armed tidal response as it moves away from the host galaxy. Gas resolution is $2000\,M_\odot$, and the gravitational softening length is 10\,pc, effectively acting as our limits to determining the mass and radii of potential clouds. This mass resolution is similar to that of the study of GMCs in simulations of \citet{2011MNRAS.417.1318D}, but less than more recent simulations of galactic discs in isolation \citep{2016MNRAS.461.1684F,2017MNRAS.464..246B} or zoom-in's of sections of whole-disc simulations \citep{2016MNRAS.458.3667D,2017MNRAS.469..383J,2018MNRAS.474.2028R}. Still, this is, to the authors knowledge, the first time a study has addressed the GMC population in simulations of tidally induced spiral arms. See P2017 for details of the simulation and discussion of the global evolution of the gaseous and stellar system. Note that our simulations take no account for magnetic fields. There is therefore an additional means of pressure support that is not included in the identified cloud population.

We also perform a separate simulation for which we allowed the disc to evolve in isolation without the perturbing companion. This simulation was evolved for a period of 600\,Myr, over which time a flocculent, many-armed spiral structure is seen. GMCs in this calculation act as a baseline for the results shown throughout the manuscript.

\subsection{GMC extraction}
\label{sec:gmcselection}
Clouds are extracted from the galaxy simulation using the ``friends-of-friends" approach of \citet{2015MNRAS.446.3608D}. A subsection of gas particles above a density threshold, $\rho_{\rm cut}$, are first extracted form the global simulation. A minimum mass of clouds is defined by the minimum particle number within a cloud, $n_{\rm pop}$. For each particle a neighbour search is performed within a scale length, $l_{\rm neigh}$, and if a particle is found it is then defined as a member of the current cloud (so long as it satisfies the density criterion). While we have three free parameters when defining clouds, in reality $\rho_{\rm cut}$ and $l_{\rm neigh}$ are highly degenerate, while $n_{\rm pop}$ is limited by resolution and the need to resolve cloud structure. We use a minimum of 40 particles to define a cloud (giving a mass resolution of $8\times 10^4\,M_\odot$). Lower than this and clouds are poorly resolved, and higher severely limits the cloud population for analysis. We then choose values of $l_{\rm neigh}=15\,{\rm pc}$, and $\rho_{\rm cut}=40\,{\rm cm^{-3}}$. While there are equally viable choices of these two parameters, we are interested in how the properties of clouds change in the aftermath of a tidal disruption, and so the quantitative determination of cloud statistics is secondary to how the global population changes.

\subsection{Cloud properties}
\begin{figure}
\includegraphics[trim = 20mm 20mm 0mm 0mm,width=90mm]{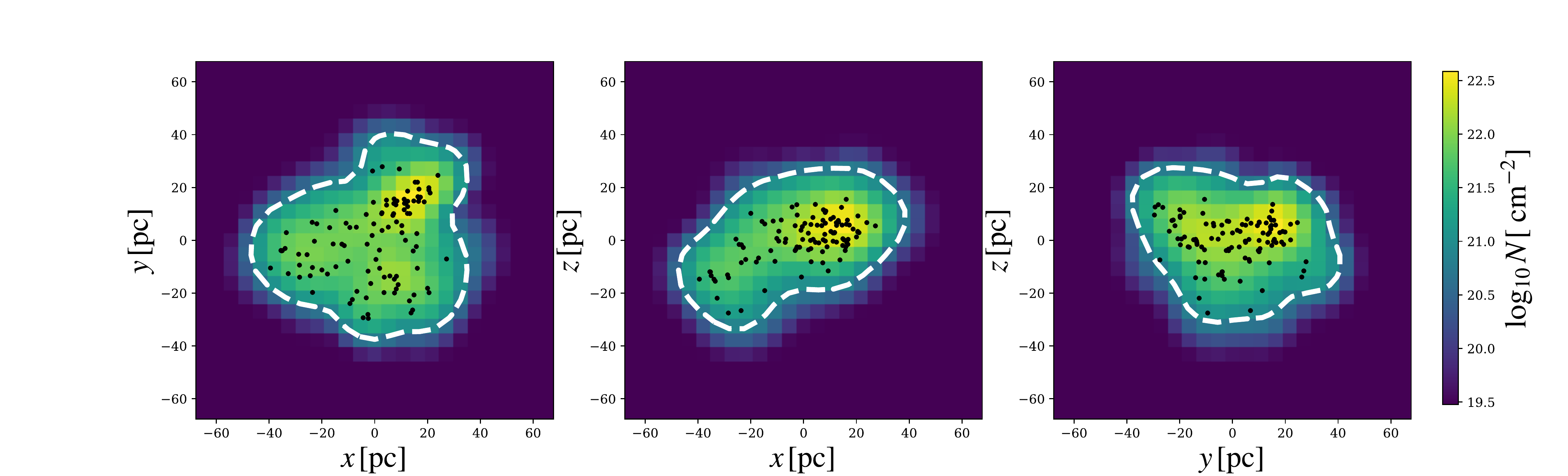}
 \caption{Example of the definition of a cloud from the galaxy simulation. The integrated surface density is shown in each plane, with black dots denoting the centre of each SPH particle. The white contour defines the surface area used for the calculation of the cloud radius.}
\label{gmc_contour}
\end{figure}

We define the mass of a cloud, $M_c$, as simply the sum of the particle masses in a given cloud. The cloud radius, $R_c$, is calculated form the mean of the surface areas in three orthogonal planes (as in \citealt{2014MNRAS.445L..65F}). These areas are defined by a contour surrounding the cloud with specific surface density. See Appendix\;\ref{Appx1} for a detailed discussion, and Figure\;\ref{gmc_contour} for an example of a GMC identified in the simulation. We define the 1D velocity dispersion, $\sigma_{\rm c}$, as:
\begin{equation}
3\sigma_{\rm c}^2 =\frac{1}{N_i}\sum_i (v_x-v_{{\rm CM},x})^2+(v_y-v_{{\rm CM},y})^2+(v_z-v_{{\rm CM},z})^2,
\end{equation}
where each velocity component of the constituent SPH particle is re-centred to the centre of mass (CM) of the entire cloud, and $N_i$ is the number of gas particles in a given cloud. We quantify the degree of how well bound a cloud is via the virial parameter, which describes the balance between gravity and internal pressure support. This is defined as:
\begin{equation}
\alpha_{\rm vir} = \frac{5\sigma_c^2R_c}{GM_c}
\end{equation}
where $\alpha_{\rm vir}=1$ defines a virialised cloud, with $\alpha_{\rm vir}=2$ used to define a marginally bound cloud (e.g. \citealt{1992ApJ...395..140B}). 

\section{Results}
\label{SecRes}

\subsection{Evolution of the galaxy}

\begin{figure*}
\includegraphics[trim = 0mm 0mm 0mm 0mm,width=170mm]{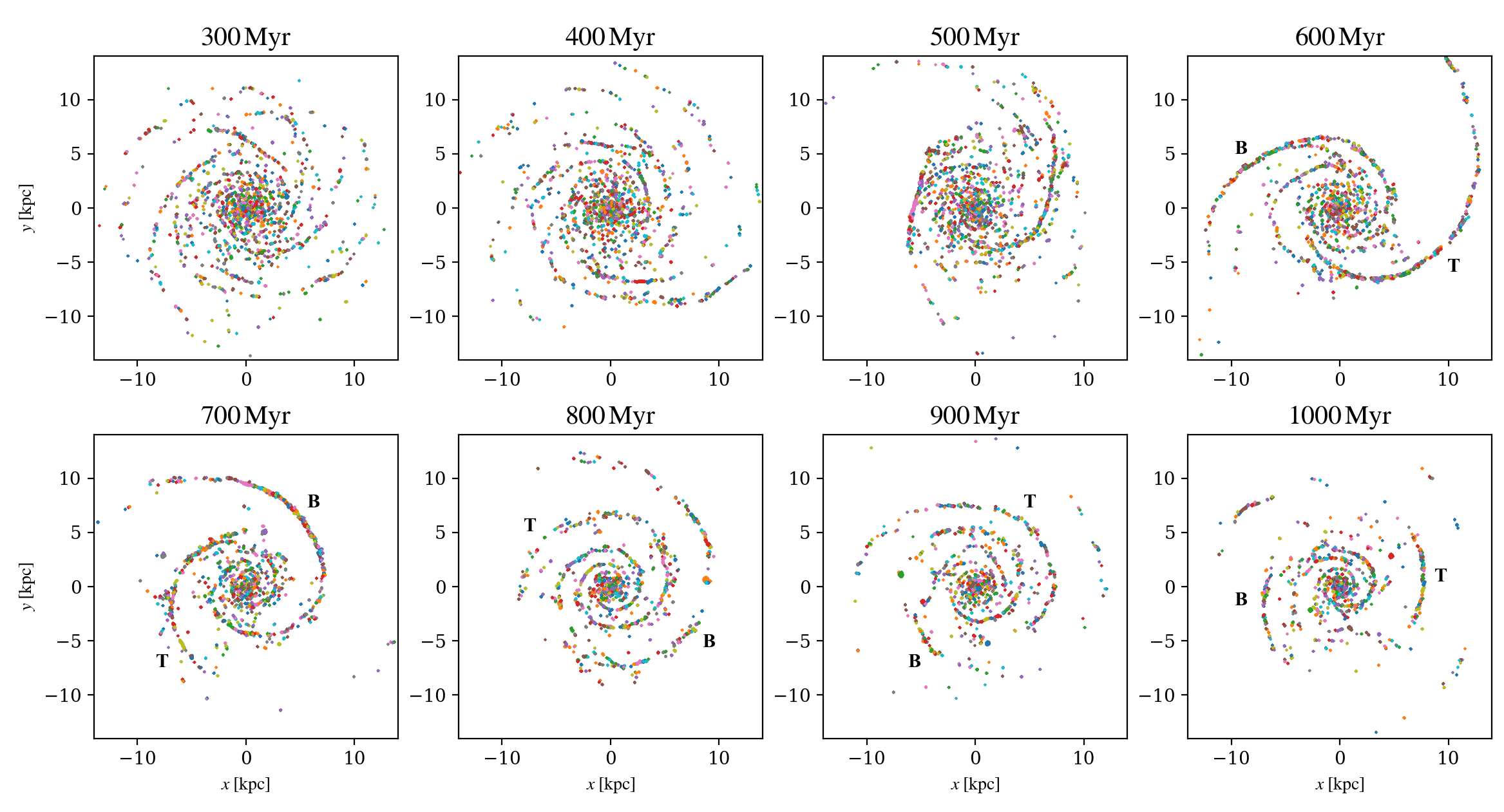}
 \caption{Gas that constitutes clouds in 8 simulation timeframes, coloured by a cloud identification number. The perturbing companion moves in clockwise from the north-east and moves away to the south-west, reaching closest approach at around 400\,Myr. The bridge and tail arms are labelled as B and T in each panel after the companion passage.}
\label{gmc_pos}
\end{figure*}

The structure and evolution of the perturbed galactic disc is described in detail in P2017, so we only give a brief overview here. The disc is initially a many-armed/flocculent disc, which has been evolved for 400\,Myr in isolation to approach a quasi-equilibrium state. The arms in this instance are driven by gravitational instabilities in the disc, referred to as \emph{dynamic} spiral arms in the literature due to their radially varying pattern speed and transient nature \citep{2014PASA...31...35D}. A companion with a mass approximately 10\% that of the primary then approaches on a prograde parabolic orbit in-plane, triggering the creation of a two-armed spiral structure, resulting in a burst of star formation. These arms then wind up on the order of a Gyr.
Figure\;\ref{gmc_pos} shows the location of extracted GMCs in 8 different times as the galaxy responds to the companion. Clouds are coloured simply by some identification number, which is not retained between time-frames, merely acting to distinguish clouds. Bridge and tail arms are labelled B and T respectively. At all times clouds appear to be highly coincident with spiral arms, be they those inherent to the disc or driven by the tidal perturbation. There is also a collection of clouds in the centre of the disc that are not strongly associated with any morphological feature and are more randomly distributed.
 
\subsection{Global cloud properties in the perturbed disc}

\begin{figure}
\includegraphics[trim = 5mm 0mm 0mm 0mm,width=90mm]{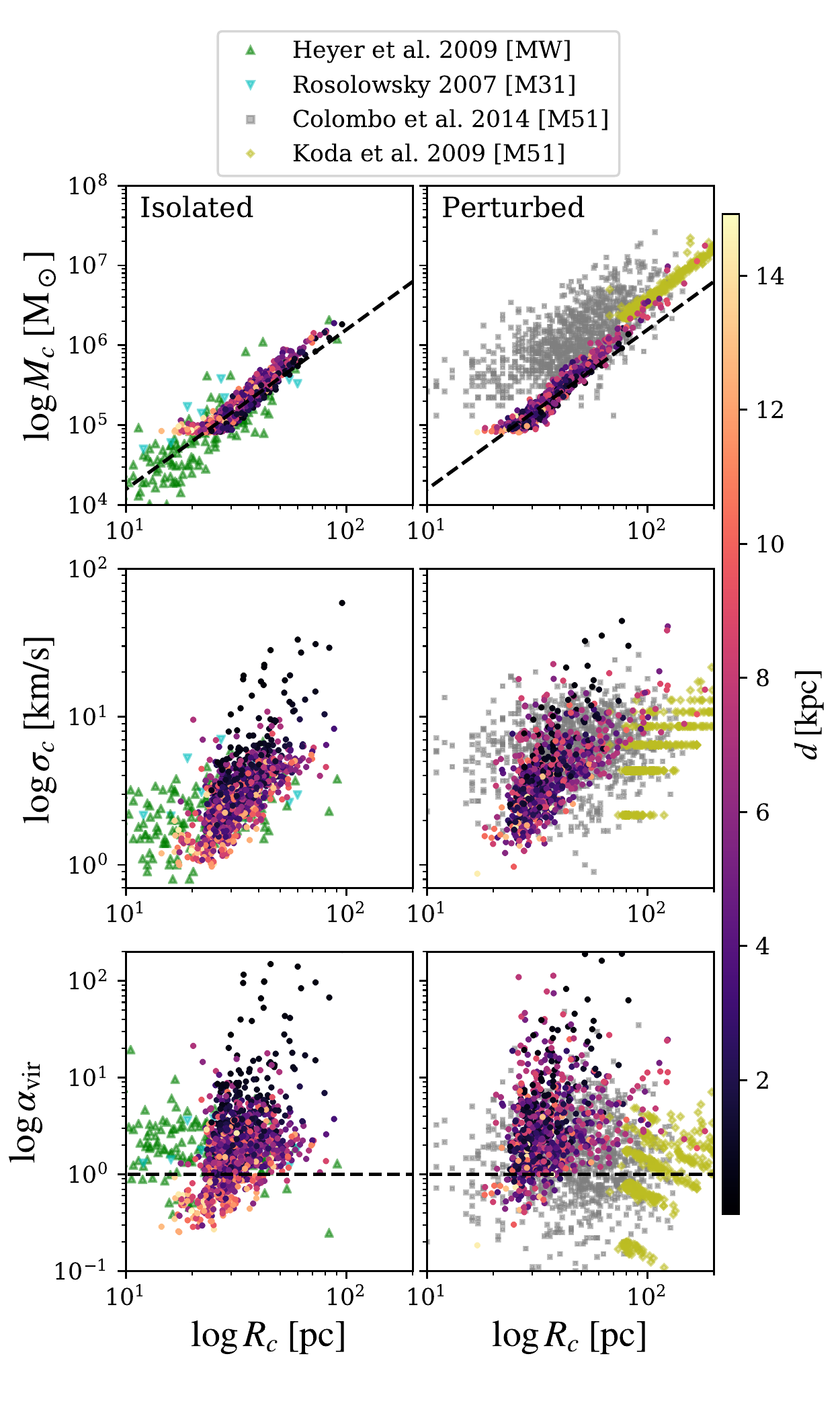}
 \caption{GMC properties in the disc when isolated (left) and post-interaction (right). The green and cyan triangles indicate the GMC data from \citet{2009ApJ...699.1092H} and \citet{2007ApJ...654..240R}, while the grey squares and yellow diamonds show the data of \citet{2014ApJ...784....3C} and \citet{2009ApJ...700L.132K}. GMCs are colour-coded by their distance from the centre-of-mass of the galactic bulge. The dashed line in the top panel indicates clouds with a fixed surface density of 50$\,M_\odot\,{\rm pc^2}$, and in the bottom panel indicates the limiting value of $\alpha_{\rm vir}=1$.}
\label{gmc-t00100}
\end{figure}

The various properties of GMCs in the isolated galaxy and the perturbed galaxy cases are shown in Figure\;\ref{gmc-t00100}. The different panels show the mass, velocity dispersion and virial parameter of the clouds, which are colour-coded by their distance from the bulge CM of the host galaxy. Observational data for the Milky Way from \citet{2009ApJ...699.1092H} and M31 from \citet{2007ApJ...654..240R} are shown as green and cyan triangles respectively for comparison with the isolated disc, and for M51 \citep{2009ApJ...700L.132K,2014ApJ...784....3C} for comparison with the perturbed case. The data for the perturbed simulation is shown at the 700\,Myr timestamp, where it shows a similar morphology to M51, and the isolated disc is shown at 600\,Myr (at the end of that simulation). Our GMCs in the isolated disc agree relatively well with the observed data for the Milky Way and M31, though do not push down into the same low mass regime for the Milky Way data due to the simulation resolution (hence the cut-off in the mass plot in the top panel at $8\times 10^4\,M_\odot$).

The data from the PAWS survey \citep{2014ApJ...784....3C} spans similar velocity dispersions and virial parameters as the simulated data, but favours somewhat higher masses than the simulation data. We attribute this to the surface density in the simulation being lower than that of M51. Mean values for surface density in the simulation in the disc are $10$--$20\,{\rm M_\odot pc^{-2}}$, while M51 shows gas surface densities in the range of $10$--$100{\rm M_\odot\,pc^{-2}}$ \citep{2009A&A...495..795H,2013ApJ...779...45M}. The GMCs of \citep{2009ApJ...700L.132K} appear to trace a different region of parameter space than the PAWS or simulated data, though this is likely due to the coarser resolution and thus a catalogue rich in massive GMAs (giant molecular associations).

The isolated and perturbed population do appear different, with the perturbed disc pushing to higher masses and velocity dispersions, thus favouring a more unbound cloud population. The significance of these differences will be discussed later in this section. A number of clouds lie on/around the stability limit in all cases, $0.5<\alpha_{\rm vir}<2$, though there are clearly many that appear unbound, which has also been noted in observational and theoretical studies \citep{2007ApJ...654..240R,2011MNRAS.413.2935D}. There is also a trend with the higher velocity dispersion clouds being at smaller galactic radii in the simulated data, resulting in a higher virial parameter closer to the galactic centre. This is likely a result of the kinematically hot stellar bulge that dominates the central mass distribution of the galaxy.

\begin{figure}
\includegraphics[trim = 0mm 0mm 0mm 0mm,width=80mm]{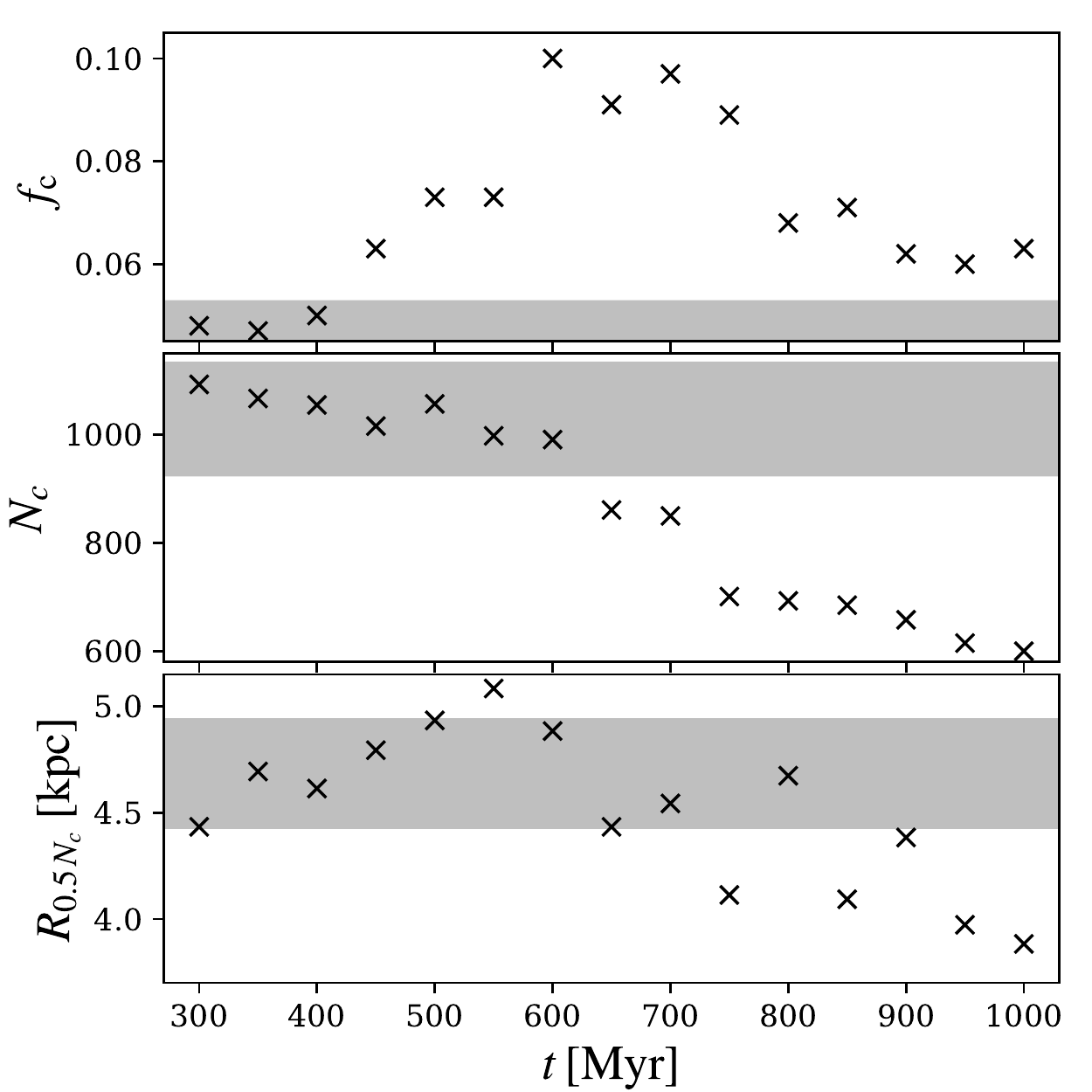}
\caption{Time series of the fraction of gas within GMCs ($f_c$, top), the number of GMCs over the course of the simulation ($N_c$, middle) and the radius that contains half of $N_c$ added cumulatively from the galactic centre ($R_{0.5\,N_c}$, bottom). Closest approach occurs at approximately 400\,Myr.}
\label{gmc-time}
\end{figure}

\begin{figure}
\includegraphics[trim = 0mm 0mm 0mm 0mm,width=90mm]{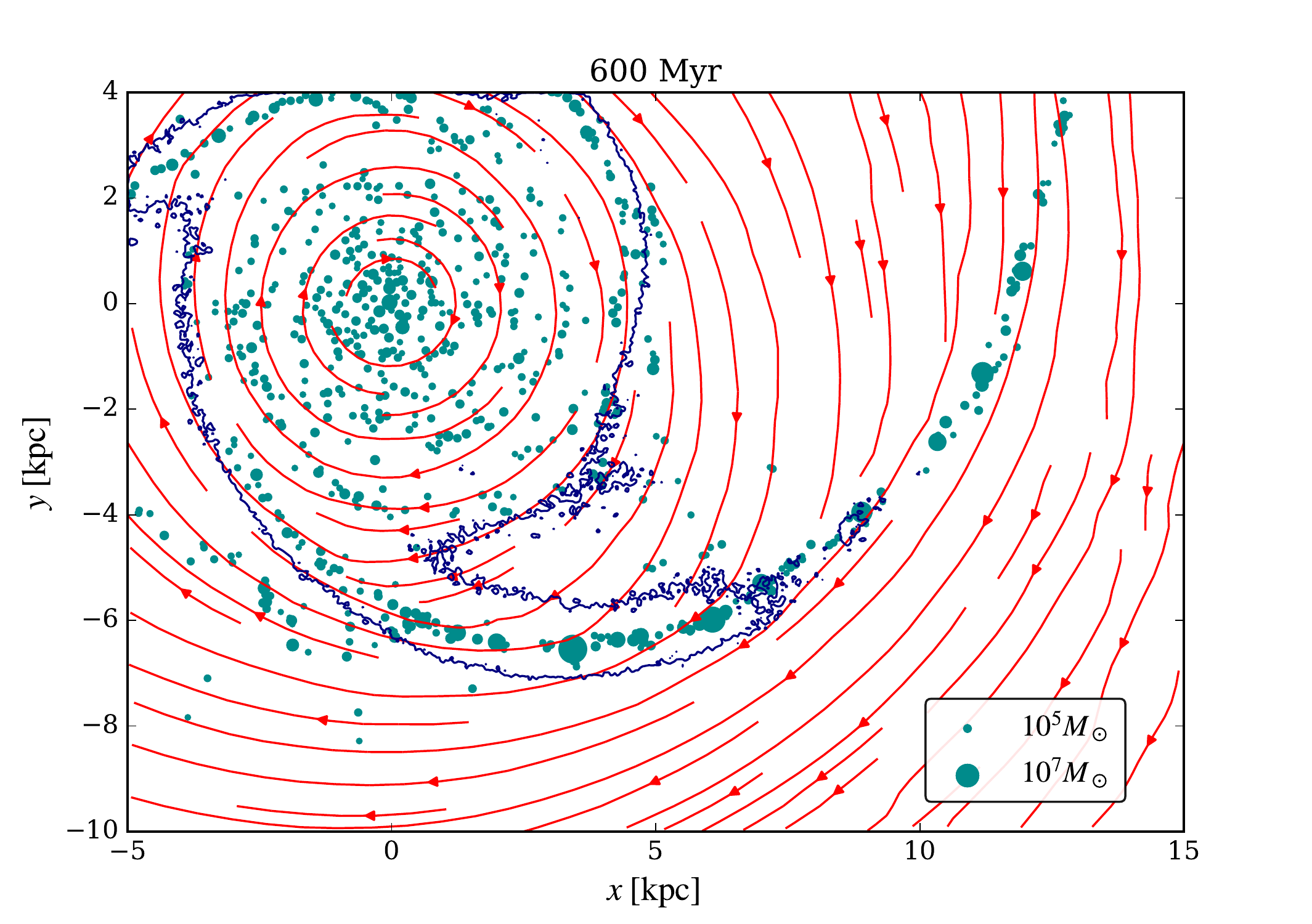}
\caption{Velocity field lines around the tail arm at 600\,Myr. GMCs are shown as cyan circles, with sizes corresponding to their masses. The blue contour illustrates the location of the stellar arms.}
\label{gmc-stream}
\end{figure}

In Figure\;\ref{gmc-time} we show the fraction of gas in the galaxy that is contained within GMCs ($f_c$, top) and the total number of clouds (middle) as a function of time. There is a clear increase in the amount of gas contained within clouds soon after perigalacticon passage (400\,Myr), which reaches a peak at 600\,Myr, when the spiral arms reach their peak strength and almost all GMCs lie within the two main arm features. This then drops down as the spiral arms begin to wind up, though does not reach pre-interaction levels in the timeframe explored here. The number of GMCs however does not increase in the same manner as $f_c$. $N_c$ stays at approximately 1000 for up to 600\,Myr, only dropping down after this period when the arms are winding up. The cloud population drops to around 700 in the later stages and appears to reach an approximate floor that coincides with the levelling off of $f_c$. In each panel of Figure\;\ref{gmc-time} we show a shaded region of parameter space. This corresponds to the range of GMC properties that is exhibited by isolated disc when evolved with no external perturbation (over a time period of 400\,Myr). The trends shown by $f_c$ and $N_c$ clearly show deviations away from the isolated population.

To understand what drives the increase in $f_c$  we show a zoom-in of the tail arm in the disc in Figure\;\ref{gmc-stream}, shortly after closest approach when the GMCs host almost 10\% of the disc gas. GMCs are shown as cyan points, sizes indicating their mass, and velocity field lines are over-plotted as red arrows, with a single contour of stellar surface density to illustrate the arm locations. Clear departures from circular rotation can be seen in the velocity field around the spiral arms, especially clear near $(+5,-6)$\,kpc. The cloud-front resides preferentially on the convex side of the arm. Streaming motions appear to drive gas tangentially along an arm once it enters for a period of time, allowing the accretion of material into large GMCs. This orbital-crowding accumulation of gaseous material is what is driving the peak in $f_c$ in Figure\;\ref{gmc-time} at 600\,Myr, with velocity fields appearing much more circularised at later times as the arms decrease in strength, causing $f_c$ to drop back down.

The bottom panel of Figure\;\ref{gmc-time} highlights a possible reason for the steady decay of $N_c$. This panel shows the galactic radius that contains half the total number of GMCs at a given time, summed from the centre outwards ($R_{0.5N_c}$). Changes in this value indicate migration of the cloud population outwards or inwards. A small increase is seen at early times due to the immediate tidal attraction. At later times the cloud population begins to move inwards, indicating a migration/inflow of dense gas regions to smaller radii and a general contraction of the gas disc. This migration would result in the merger of clouds in the inner disc, acting to lower $N_c$ without necessarily changing $f_c$, and Figure\;\ref{gmc-time} does show a correspondence between the drop in $N_c$ and $R_{0.5N_c}$ after 600\,Myr. Non-axisymmetric features have been known to induce radial migration such as this in simulations of galaxies \citep{2002MNRAS.336..785S,2016MNRAS.458.3990P,2016MNRAS.461.1684F}. The isolated disc values of $R_{0.5N_c}$ vary by about 0.5\,kpc, which almost encompasses the peak at 550\,Myr in the perturbed disc properties, but is well out of range of the drop at later times while the spiral arms are winding up.

\begin{figure*}
\includegraphics[trim = 10mm 0mm 0mm 0mm,width=180mm]{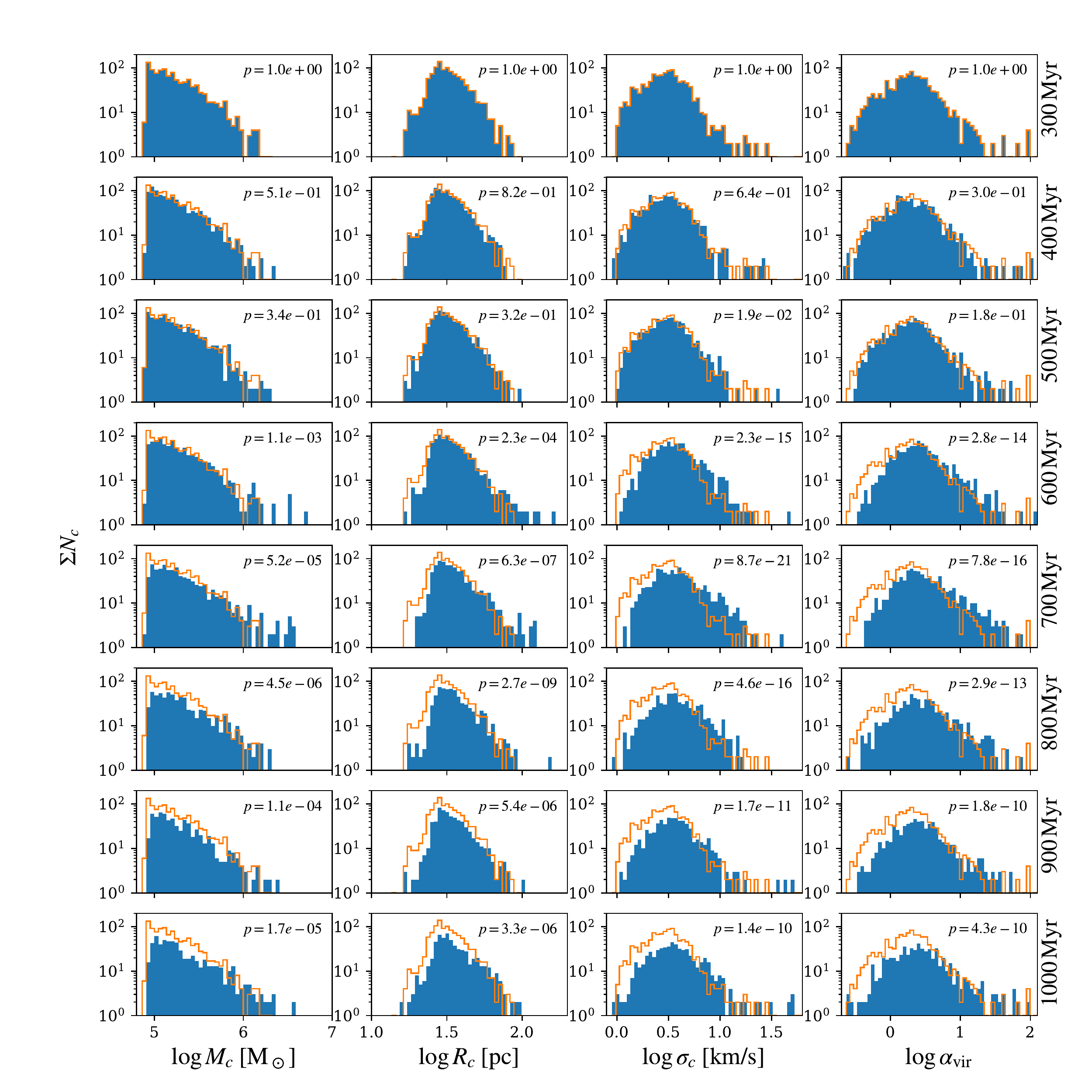}
\caption{Histograms of GMC radii, virial parameter, masses and velocity dispersion as a function of time (increasing downwards). The orange outline indicates the properties at 300\,Myr when the system is effectively in isolation. In the top right is the $p-$value for the Kolmogorov-Smirnov test applied to the distribution of each GMC property compared to that at 300\,Myr.}
\label{gmc-timeprops}
\end{figure*}

To further hone-in on the evolution of the clouds we show a number of histograms of cloud properties in Figure\;\ref{gmc-timeprops}. Each column shows histograms of $M_c$, $R_c$, $\alpha_{\rm vir}$ and $\sigma_c$, with each row showing a different time-frame. The properties of the 300\,Myr snapshot, when the system is effectively in isolation, is repeated as the orange outline in each panel. The properties between the 300--400\,Myr timestamps are effectively the same, confirming the system is in equilibrium in terms of cloud properties when in isolation. After perigalacticon passage there is then a shift in properties of the clouds. The mass of clouds tends to push up into the high mass regime, and reduce the lower mass population over the course of the simulation. The radii of the clouds also increases over time, with a steady decrease in the quantity of smaller clouds. The velocity dispersion of the clouds shows the strongest change after the interaction, with the population of clouds with only small velocity dispersions ($\sigma \leq$4\,\kms{}) being severely reduced. These changes result in clouds that appear to be bound much weaker than in the isolated case, seen in reduction of clouds with low values of $\alpha_{\rm vir}$. 

To test for the statistical significance of the properties at different times, we perform Kolmogorov-Smirnov tests\footnote{This tests whether two samples are drawn from the same distribution using differences in cumulative distributions, with the $p$--value indicating the significance to which you can reject the null-hypothesis that the samples are statistically different.} on each dataset in Figure\;\ref{gmc-timeprops} comparing each histogram to the data at the 300\,Myr timeframe when the system is in isolation. $p$--values for this test are given in the top-right of each figure. As expected, the 400\,Myr timeframe appears to be drawn from the same distribution as that at 300\,Myr, as one would hope given the system remains unperturbed at this time. As the system evolves the $p-$values lower dramatically, indicating statistically significant changes in the GMC properties, especially the distributions of $\alpha_{\rm vir}$ and $\sigma$. Similar histograms were made for the GMCs of the isolated disc, and show only minor changes between time-frames. This is supported by corresponding KS-tests, which had $p$--values of 0.4--0.9 between different time-frames for all four measured parameters.

The net result of the interaction therefore appears to generate larger and more massive clouds structures along tidal arms. However, the tidal perturbation induces large non-circular motions in the gas, which are manifest by higher velocity dispersions and clouds that are less well-bound than their brethren in the isolated galaxy.

\subsection{Changes in the cloud mass function}

\begin{figure}
\begin{centering}
\includegraphics[trim = 10mm 0mm 0mm 0mm,width=70mm]{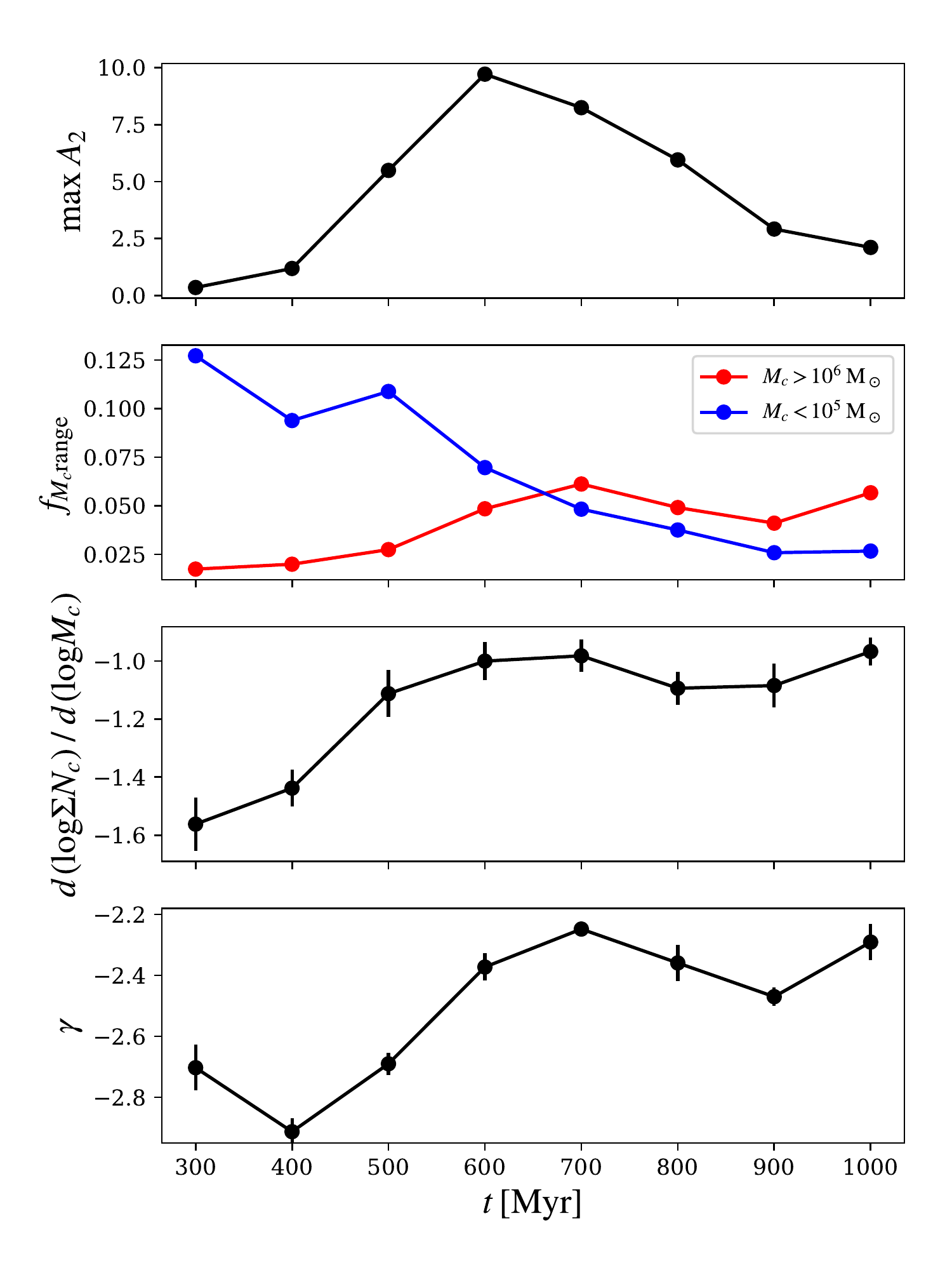}
\caption{Plot of various diagnostics of GMC and spiral arms over time. Top panel shows the maximum power of the $A_2$ mode in the old spiral arms. Second panel shows the fraction of high mass ($>10^6\,{\rm M_\odot}$) and low mass ($<10^5\,{\rm M_\odot}$) clouds. The third panel shows the gradient of the slope of the mass function of clouds (left panels of Fig.\;\ref{gmc-timeprops}), while the bottom panel shows the slope of the cdf of the mass function for $M_c>10^{5.5}\,{\rm M_\odot}$. Uncertainties in fits were estimated from bootstrap sampling and taking the standard deviation of the fitted gradients.}
\label{gmc-fourier}
\end{centering}
\end{figure}

It is important to better understand how the cloud mass function is changing, as presented in the previous section. In Figure\;\ref{gmc-fourier} we show some diagnostics of this change in GMCs. In the top panel we show the strength of the spiral arms over time. This is calculated from the peak power of the $A_2$ component of the Fourier decomposition over $2{\rm \,kpc}<R<10{\rm \,kpc}$ for the entire ``old" (present at $t$=0\,Myr) stellar population, which is an often-used proxy for arm and bar strength. As expected, the amplitude of the arms rises fast and slowly decays after the closest approach of the companion. In the second-from top panel we show the fraction of clouds in a certain mass range. The blue line shows the fraction of low mass clouds ($<10^5\,{\rm M_\odot}$) and the red line shows the population of high mass clouds ($>10^6\,{\rm M_\odot}$). Clearly there is a preference for an increase in the population of high mass clouds, and the destruction of low mass clouds as a result of the spiral perturbation. Low-mass clouds seem to continually disappear, even as the spiral dissipates, whereas high-mass clouds experience an initial boost in number and then level off once the spiral starts to wind up and disappear. This continued reduction in low mass clouds is likely due to the spiral arms continually sweeping up smaller clouds into larger complexes. Despite the spiral dissipating at later times, there is a still a large degree of structure in armlets, branches and filaments, which have wound up much tighter than the initial strong spiral. As such, smaller clouds in the inter-arms quickly encounter over dense regions after leaving an arm, hindering the ability of the disc to maintain a low mass cloud population.

It should be noted that the smallest clouds are on the threshold of our identification routine. As such it could be possible that slight dynamical changes in the galaxy may induce a large difference in this population of borderline-defined clouds due to our resolution, which may play a part in accentuating trends in the low-mass cloud population.

In the third panel of Figure\;\ref{gmc-fourier} we show the gradient of the mass spectra that are shown in the left column of Figure\;\ref{gmc-timeprops}, using only the moderate to high mass regime ($>10^5\,{\rm M_\odot}$) where the slope of the population appears near-linear in log-space. We fit a simple 1$^{\rm st}$ order polynomial to the slope in this regime, and plot the gradient as a function of time. Lower values of this gradient (more negative) indicate a steeper mass function and a higher relative fraction of low mass clouds. The slope can clearly be seen to flatten as the spiral arms increase in strength, though maintains the same level once the spiral begins to wind up, indicating the cloud population is irrevocably changed after the tidal perturbation. It is expected that the cloud population would return to the original distribution eventually, though we only ran our calculation for 1.2\,Gyr at which point there were still clear residual features caused by the interaction. While grand design 2-armed spiral structures tend to dissipate after about 1\,Gyr \citep{2015ApJ...807...73O,2016MNRAS.458.3990P}, gas migration and the reduced global gas reservoir caused by the boost in star formation may mean the GMC population will never truly return to that pre-interaction. Other simulations in the literature ran for much longer time-periods also show discs displaying altered morphologies many Gyr after such an interaction (e.g. \citealt{2011MNRAS.414.2498S,2018MNRAS.474.5645P}).

In the bottom panel we show the gradient of the cumulative mass spectra, which is often measured by observational works. We fit the truncated power law distribution of the form:
\begin{equation}
N(M'>M_c)=N_0 \left[      \left(  \frac{M_c}{M_0}\right)^{\gamma+1}  - 1   \right],
\label{massCDF}
\end{equation}
for clouds below some mass $M'$. $\gamma$ is the slope of the mass function, with higher values indicating dense gas resides more in high mass clouds, and lower values indicating a preference for a lower mass population \citep{1997ApJ...476..166W}. We fit this function to our GMC populations, keeping $N_0$ and $M_0$ as free parameters\footnote{We use the Nelder-Mead simplex algorithm in \textsc{python}'s \textsc{scipy} package and a $\chi^2$ minimising statistic. The results of this fitting procedure are given in Appendix\;\ref{Appx2}.}; see \citet{2005PASP..117.1403R} for a discussion of their significance. We use the data in the mass range of $\log M_c>5.5$ for this fitting procedure, with lower cloud masses nearing our resolution limit and higher values requiring a fit to a substantially limited cloud population (the order of 10s of clouds) making the fit quite susceptible to a single evolving massive cloud. The values of $\gamma$ in Figure\;\ref{gmc-fourier} confirm that of the previous panel, in that the cloud mass spectrum moves up to higher masses after the passage of the companion. The values initially start quite low, similar to values of the inter-arm regions of M51 from \citet{2014ApJ...784....3C} who fit their GMC mass spectra to Equation\;\ref{massCDF} in different regions of the disc. M33 also shows low values of $\gamma$ in \citet{2005PASP..117.1403R}, who reported a value of $-2.9 \pm 0.4$, though other studies have reported higher values closer to $-2.0$ \citep{2007ApJ...661..830R,2012A&A...542A.108G}. Interestingly, \citet{2011ApJS..197...16W} see large variations in values of $\gamma$ for the same dataset of clouds in the LMC, depending on how the clouds are defined and how the disc is decomposed. This seems a recurring theme, with many observations of the same galaxy offering quite different values for $\gamma$, questioning its robustness as a metric of a given cloud population (see the discussion in \citealt{2017ApJ...841...43E}). Note that when the disc is in isolation it exhibits material arm like structures with a steep power law slope compared to the more wave-like arm features after isolation. This is also seen in the decomposition of the M51 disc by \citet{2014ApJ...784....3C} who separate arm regions in material and density wave arm components.

\subsection{Cloud properties in different environments}

\subsubsection{Variation of cloud properties with location}

\begin{figure*}
\includegraphics[trim = 10mm 0mm 20mm 0mm,width=140mm]{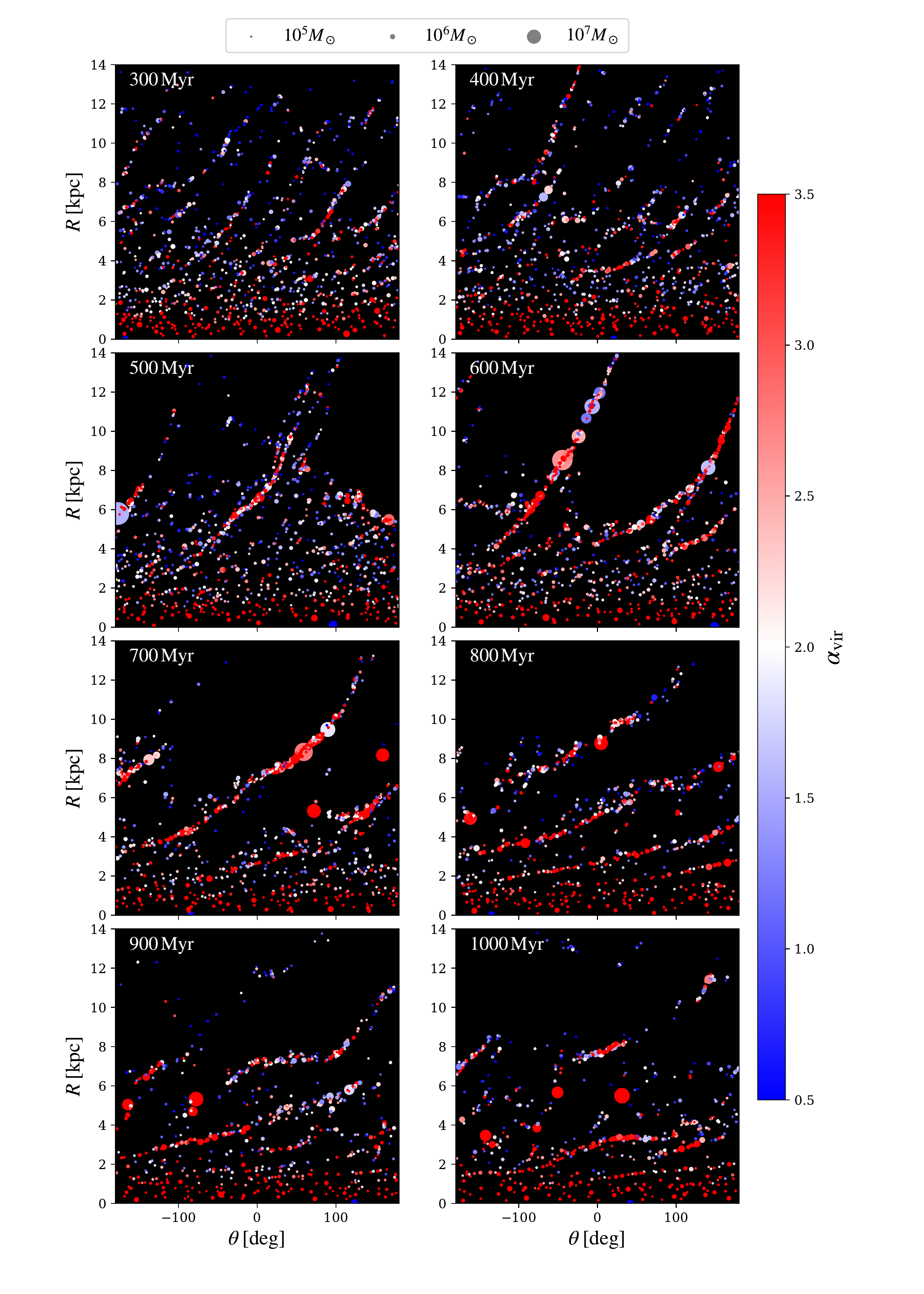}
\caption{Cloud positions in eight 100\,Myr timeframes in radius-azimuthal position. Points are coloured by $\alpha_{\rm vir}$ parameter, centred on $\alpha_{\rm vir}=2$, with sizes linearly proportional to $M_c$.}
\label{alphamap}
\end{figure*}

In Figure\;\ref{alphamap} we show the positions of clouds at eight different timestamps in the simulation, plotted in radius-azimuth space in the disc. Clouds are coloured by $\alpha_{\rm vir}$, with a value of 2 indicated by the white points, red being highly unbound and blue indicating bound clouds. The sizes of the clouds indicate their relative mass on a linear scale. Pre-interaction there is a significant quantity of well bound clouds, mostly in the mid-outer disc. The spiral arms host a population of unbound clouds at this time, with the inter-arm regions dominated by the highly bound clouds. Clouds within the potential of a dynamic spiral pattern are likely still in an early stage of their lives, having not yet achieved a bound state. Once the arm dissipates these clouds are sheared away or shed mass via feedback until only bound cores remain, leading the inter-arms to be dominated by bound clouds (see also Sec. \ref{sec:tracker}). The clouds in the inner disc appear highly unbound. This is the bulge-dominated region, which supplies an extra component of velocity dispersion to the clouds, keeping them from remaining bound for long periods. Similar plots of the isolated simulation over 600\,Myr show no discernible changes from the cloud population shown in the top left of Figure\;\ref{alphamap}.

At 600\,Myr the tidal interaction has created a strong 2-arm spiral arm pattern, a bridge arm pointing to the perturber, and a tail arm on the other side of the disc. Clouds are now highly coincident with the spirals in the mid/outer disc. The arms predominantly contain numerous unbound clouds, with arm clouds reaching much higher masses than in the isolated disc. There still exists a population of well-bound inter-arm clouds, seen clearest on the concave side (upstream) of the bridge arm ($|\theta|>120^\circ$, $R>4$\,kpc), seen also in Fig. \ref{gmc_pos}. The tail arm (centred around $-40^\circ$) contains a population of bound clouds beyond 10\,kpc, which is not as evident as bridge arm or at smaller radii. This coincides with the accumulation of dense gas and burst-like star formation episode discussed in P2017 that is seen in the tail but not the bridge arm at that given time. Differences between cloud populations from arm to arm in bisymmetric spiral galaxies could thus act as a diagnostic of spiral structure generation in high resolution observations, as they could hint at a tidal progenitor seeding the spiral features.

At 800\,Myr the spiral arms have wound up considerably, and there seems to be a greater amount of well-bound clouds in the arms, though still only a few inter-arm clouds. Overly massive, loosely bound clouds appear less numerous than immediately after the interaction (i.e. at 600/700\,Myr), though still more so than in the isolated disc case. The spiral arms triggered by the interaction continue to offer safe-havens for loosely bound clouds, several galactic rotations after the companion's closest approach.

\begin{figure}
\begin{centering}
\includegraphics[trim = 0mm 0mm 0mm 0mm,width=80mm]{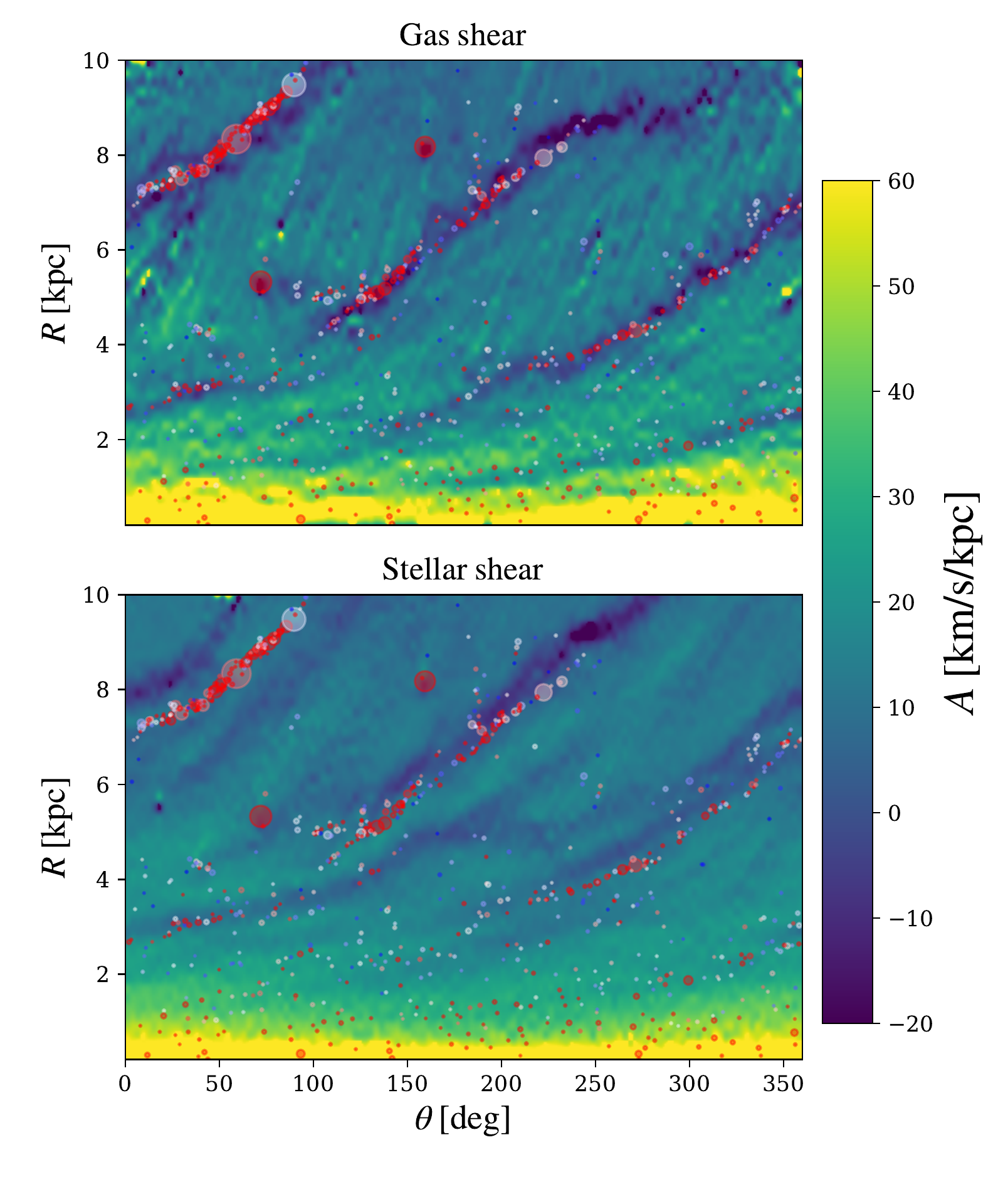}
\caption{Maps of galactic shear as a function of radial and azimuthal position after 700\,Myr of evolution. The top panel shows a map of shear, $A$, in the gas and the bottom panel shows $A$ for stars. GMCs are plotted with the same colour/size scheme as used in Figure\;\ref{alphamap}.}
\label{shear}
\end{centering}
\end{figure}

We quantify the shear in the disc as a function of position to highlight its role in creating/destroying clouds. We use Oort's $A$ constant as a measure of shear in the galactic disc, where $A$ is given by:
\begin{equation}
A=\frac{1}{2}\left(  \frac{V_c}{R}-\frac{dV_c}{dR} \right)=-\frac{R}{2}\frac{d\Omega}{d R}
\end{equation}
where $\Omega$ is the angular velocity of disc material and $V_c$ the circular velocity. Figure\;\ref{shear} shows maps of $A$ in the gas (top) and stars (bottom) for the 700\,Myr timeframe when the arms are still strong and clearly disconnected from the companion. We over-plot the locations of GMCs using the same colour and size scheme as in Figure\;\ref{alphamap}. Values of shear agree very well with those calculated by \citet{2014PASJ...66...36M} for M51 (their Fig.\;14), with negative values tracing spiral arms and higher positive values in the inter-arm regions. More significantly, we see a clear trend in low shear regions correlating with the location of GMCs, with higher shear ($A>20$\,km/s/kpc) in the inter-arm regions. This is especially clear in the gas, where shear reaches very high values in the inter-arms, and the low shearing regions lie almost exactly along the GMC over-densities. Some GMCs do exist within the high shear regions, these include some of lowest mass and most well bound clouds, with the large complexes of massive GMCs in the arms being too loosely bound to survive passage into these high shear regimes. This is in agreement with \citet{2014PASJ...66...36M} who, using the GMC data of \citet{2009ApJ...700L.132K}, find that the largest cloud complexes (GMAs) can only exist in the low shear arms and the higher shear regions are populated with much smaller clouds. \citet{2015ApJ...806...72M} also infer that shear in M51 is the primary driver of cloud death in the inter-arm regions, with feedback playing a larger role in the outer disc. While the stellar shear in the simulation is in general agreement with the gas shear, there is a weaker agreement correlation between the low shear regions and the GMC population. In particular, there are a few regions of low stellar shear with no associated GMCs, as well as a clear offset in all regions (see P2017 for a detailed discussion of offsets in this simulation).

It should be noted that are two massive, clearly unbound clouds in the inter-arm regions in Figure\;\ref{shear}. These clouds appear to be bound primarily by stellar material, which has formed a self-gravitating clump that survives spiral arm passage. As such, the value of the virial parameter does not fully describe the dynamical state of such gas, even though it does continue to form stars. 

\begin{figure}
\begin{centering}
\includegraphics[trim = 10mm 0mm 0mm 0mm,width=95mm]{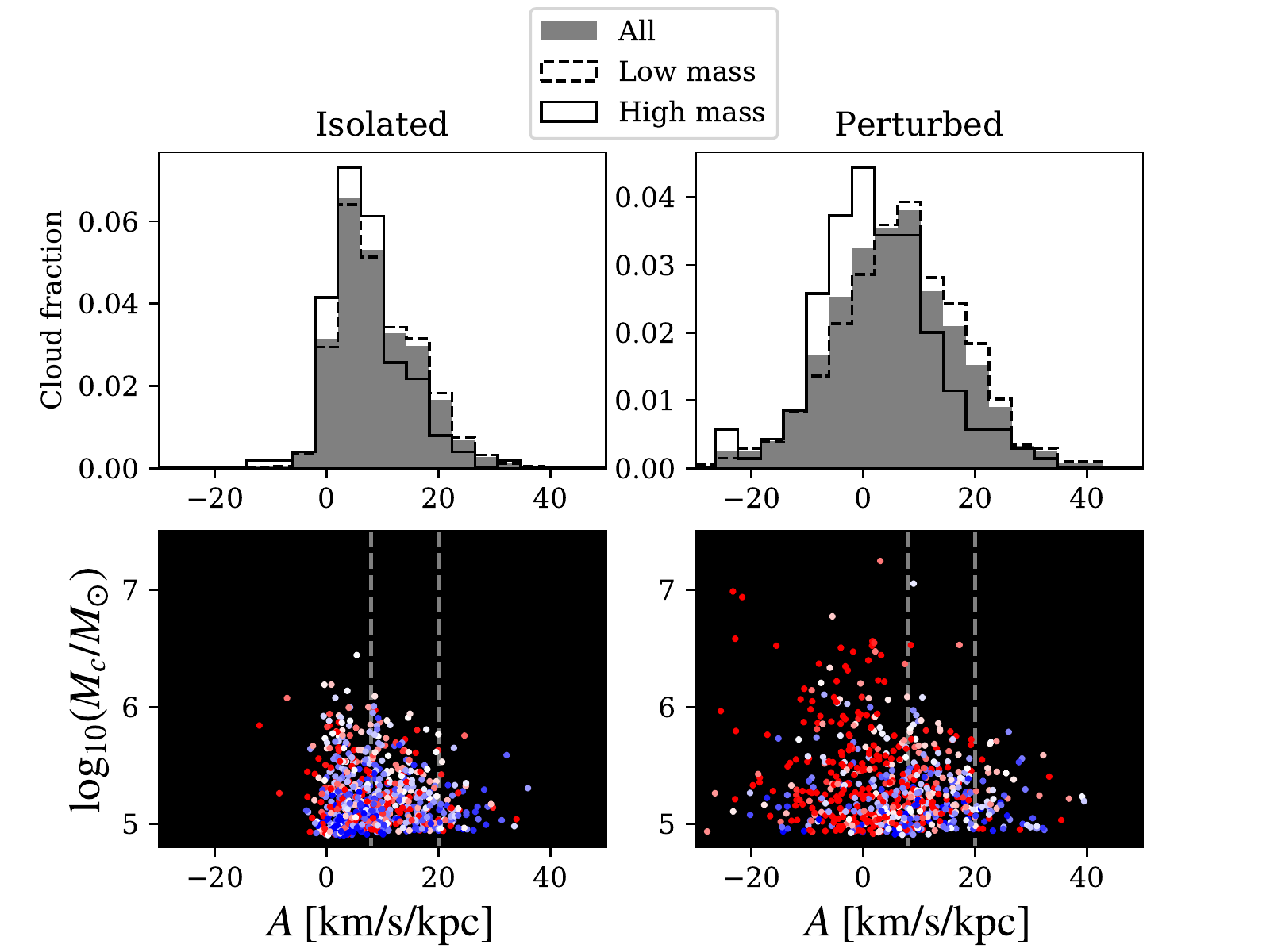}
\caption{ The relationship between GMC mass and local shear for the isolated disc (left) and the perturbed disc (right, the same data as Fig. \ref{shear}). In the lower panels the mass of clouds verses the local gas shear is plotted for all GMCs outside of $R>2$\,kpc, with points coloured by virial parameter, using the same colour scheme as Figure\;\ref{alphamap}. Dashed lines indicate the range of shear in the interarm regions. The top panel shows clouds binned by shear, but separated into low and high mass cloud populations (split by $\log{M_c/M_\odot}=5.5$).}
\label{Bishear}
\end{centering}
\end{figure}

To better understand the role of shear, we show how GMC properties directly relate to shear in Figure\;\ref{Bishear}. The lower panels show $A$ plotted against cloud mass, with GMCs coloured by the viral parameter (using the same colour scheme as Fig.\;\ref{alphamap}). We show clouds at galactic radii $>2$\,kpc to highlight the impact of shear in arm/interarm regions. In the top panels the GMCs are binned by local shear, with filled bins showing the GMCs of all masses. The solid outlined histogram shows high mass clouds ($M_c>10^{5.5}\,M_\odot$), and dashed histograms show only lower mass clouds ($M_c<10^{5.5}\,M_\odot$). Left and right columns show the GMC population in the isolated disc, and perturbed disc respectively. The isolated GMCs show a population of clouds that are distributed across a range of shears, with both high and low mass clouds being distributed across similar levels of shear. However, the perturbed disc shows some clear differences. The strong asymmetric structure creates more regions of low shear, which is associated with the higher cloud masses. The high mass GMCs in this region are clearly more loosely bound than the clouds in higher shear regions, which are instead dominated by less massive and more well bound clouds. The binned data in the upper right panel clearly shows that the high mass cloud population preferentially inhabits the low shear regime, with the data centred about $A=0$\,\ps{}. The lower mass cloud population peaks around $A=8$\,\ps{}, similar to the clouds in the isolated disc.

\begin{figure}
\begin{centering}
\includegraphics[trim = 0mm 0mm 0mm 0mm,width=60mm]{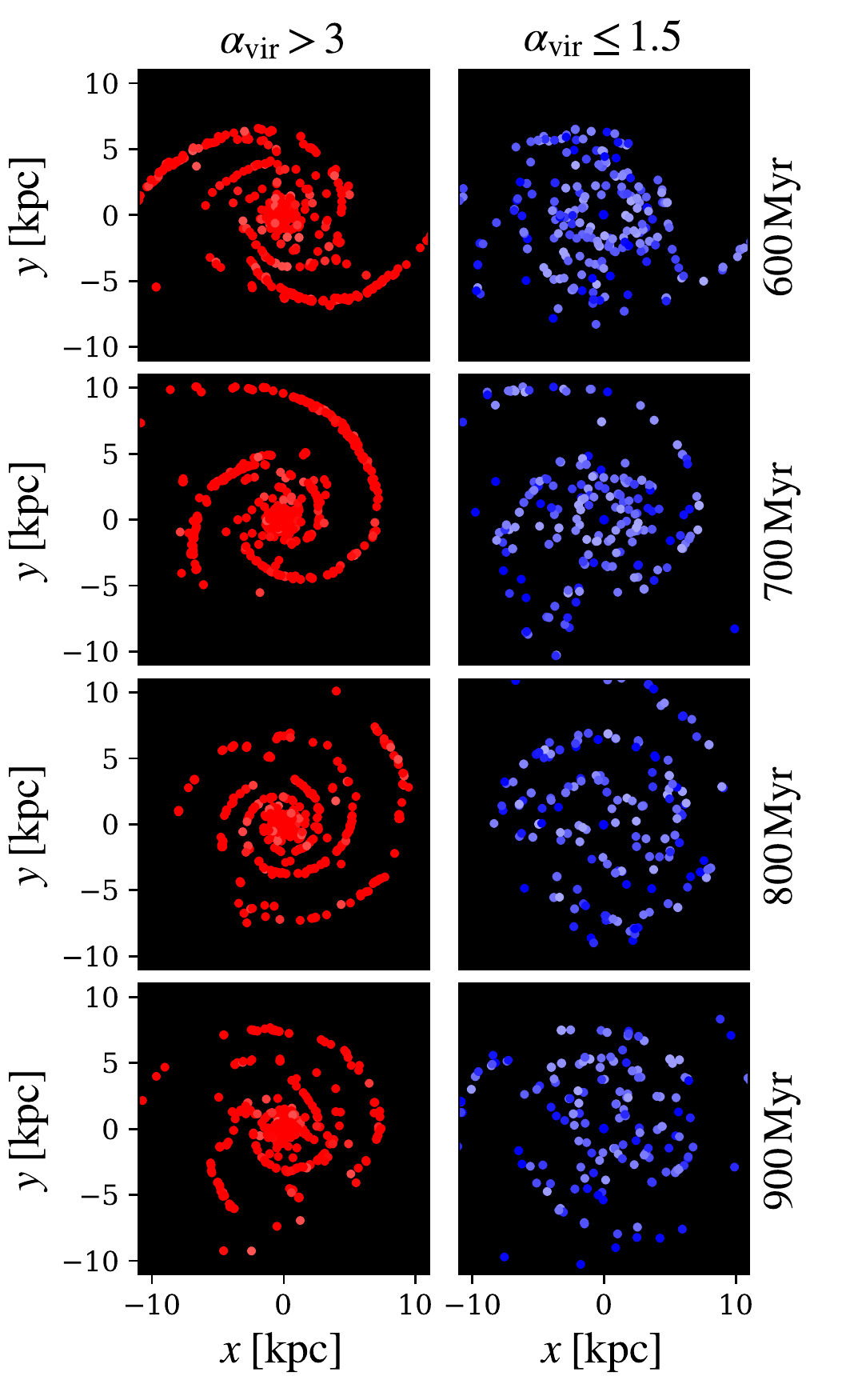}
\caption{Locations of unbound ($\alpha_{\rm vir}>3$, left) and well-bound ($\alpha_{\rm vir}\leq 1.5$, right) clouds during peak spiral strength. Points are coloured using same scale as Fig\;\ref{alphamap}.}
\label{boundness}
\end{centering}
\end{figure}

In Figure\;\ref{boundness} we show the locations of only unbound ($\alpha_{\rm vir}>3$, left) and well-bound ($\alpha_{\rm vir}\leq 1.5$, right) clouds, immediately during and after peak spiral response. There is a clear correlation between unbound clouds and the spiral arms. However, the bound clouds tend to show a much weaker correspondence with the spiral arms, with many appearing in the inter-arm regions. By tracking the history of the gas in some of these well-bound clouds we observe that while some of these inter-arm clouds appear to be the remnants of the large unbound arm complexes (e.g. Fig.\;\ref{cloudtrack1} and Appendix\;\ref{Appx3}), others seemed to have been formed in-situ from small over-densities in filaments and arm spurs. A detailed cloud-tracking analysis seems warranted to further study the individual cloud evolution, but is beyond the scope of this work.

\subsubsection{A closer look at a number of select clouds}
\label{sec:tracker}

\begin{figure*}
\begin{centering}
\includegraphics[trim = 10mm 10mm 20mm 0mm,width=180mm]{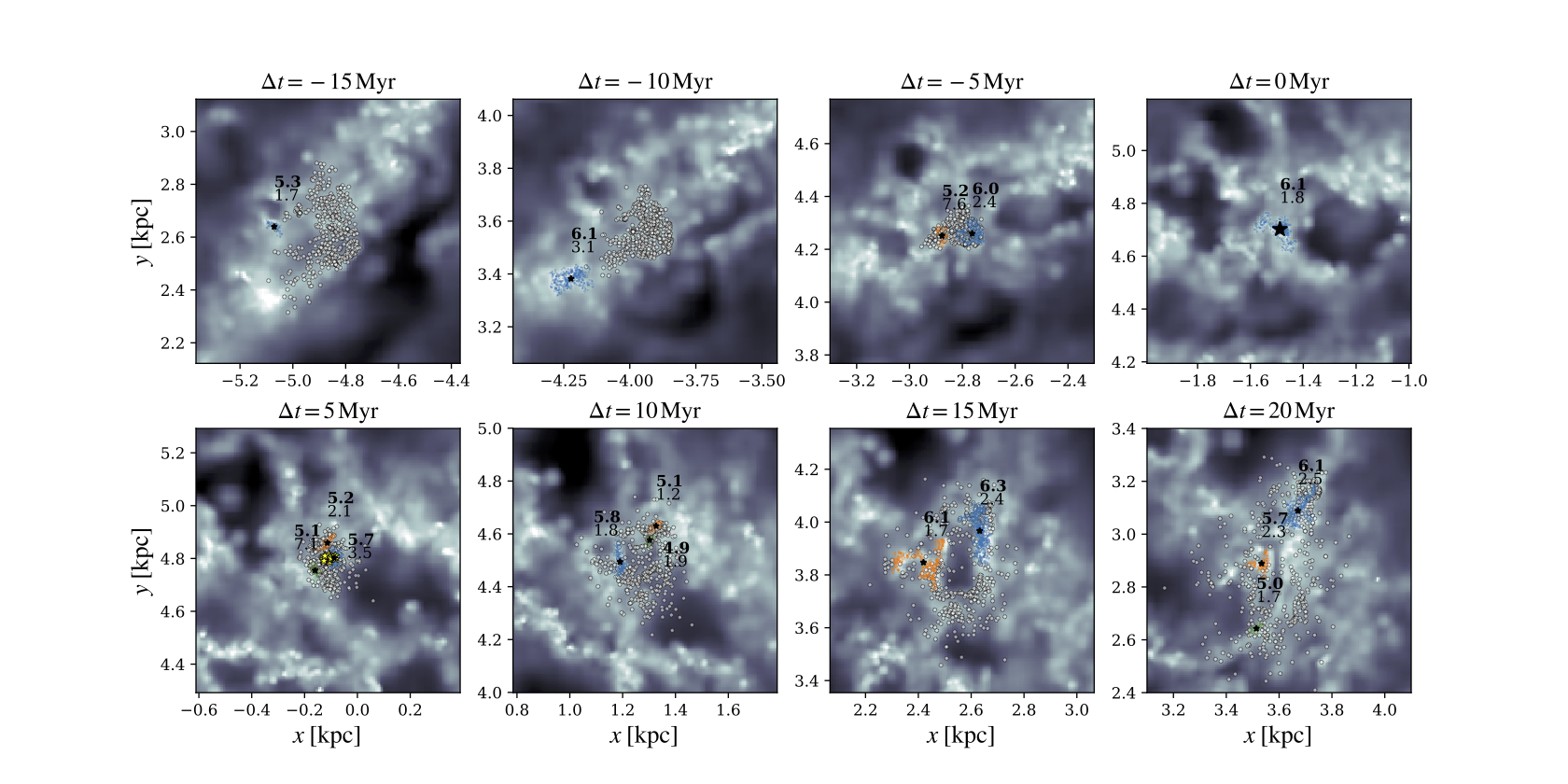}
\caption{Plot showing the evolution of a cloud and its constituent gas particles. The target cloud (Cloud1) is defined at $\Delta t=0$\,Myr by the blue particles. Coloured points shows any GMCs that contain any number of particles that were or will be part of this cloud at $\Delta t=0$\,Myr, i.e. the donor and inheritor clouds. White circles indicate the history and evolution of the actual gas particles that define Cloud1. Yellow stars indicate stars formed in Cloud1 5\,Myr after the time of definition. Black stars show the centre of mass of any cloud shown, with the numbers above indicating the values of $\log M_c/M_\odot$ (bold) and $\alpha_{\rm vir}$ for the cloud at that time. Cloud1 is mid-way on a primary arm at 700\,Myr, and appears to show the interplay between two/three distinct gas complexes, pushed apart by feedback and eventually drifting apart.}
\label{cloudtrack1}
\end{centering}
\end{figure*}

To better understand the cloud evolution we track several of the larger clouds in our sample as they evolve throughout the the disc. Similar analyses were performed by \citet{2013MNRAS.432..653D} for density-wave like spirals and \citet{2017MNRAS.464..246B} for dynamic spirals. In Figures\;\ref{cloudtrack1} and \ref{cloudtrack2} to \ref{cloudtrack9} we show top-down maps of the gas in the perturbed disc as it evolves into a GMC. The coloured background indicates gas surface density. Coloured markers show any GMC that contains gas that will or did constitute the targeted GMC (donor and inheritor clouds), which is defined in the fourth time frame ($\Delta t=0$, the only cloud shown in this panel). Above each cloud we indicate the virial parameter and mass, $\alpha_{\rm vir}$ and $\log_{10}(M_c/M_\odot)$, with the later indicated by bold font. In the panel immediately after $\Delta t=0$\,Myr we show the newly formed stars in the disc from the target cloud as yellow starred symbols. White circles indicate the positions of gas particles that make up the target cloud at $\Delta t=0$. Black starred symbols show the position of the centre of mass of each cloud. 

The only selection criteria for these tracked clouds is that they are in the higher mass range; $M_c>10^{5.8}M_\odot$ and exist at 700\,Myr. We have shown only a single example of evolution here and show additional examples of clouds in Appendix\;\ref{Appx3}.

Cloud1 in Figure\;\ref{cloudtrack1} is located on one of the two primary arms, around the mid-disc radially where inter-arm densities are moderate compared to the outer disc and there appears a decent survival rate of GMCs as they leave the spiral. A cloud can be seen that approaches a large gas reservoir at $\Delta t = -10$\,Myr, which triggers the formation of a smaller cloud at $-5$\,Myr, which then combine to a single, well bound cloud (though still not virialised). This cloud is quite small in size, and the resulting clustered star formation appears to split the cloud into three smaller objects. These resultant clouds accumulate much of the surrounding gas, though it appears that towards the last timestamp the clouds are slowly succumbing to shear and drifting apart as they move into the inter-arm region.

The remaining clouds shown in Figures  \ref{cloudtrack2} to \ref{cloudtrack9} illustrate a remarkable diversity of cloud evolution, some forming in isolation, some from clear 2-cloud interactions. The reader is encouraged to turn to Appendix\;\ref{Appx3} if they are interested in the detailed cloud evolutionary histories.

\begin{figure}
\begin{centering}
\includegraphics[trim = 10mm 0mm 10mm 0mm,width=80mm]{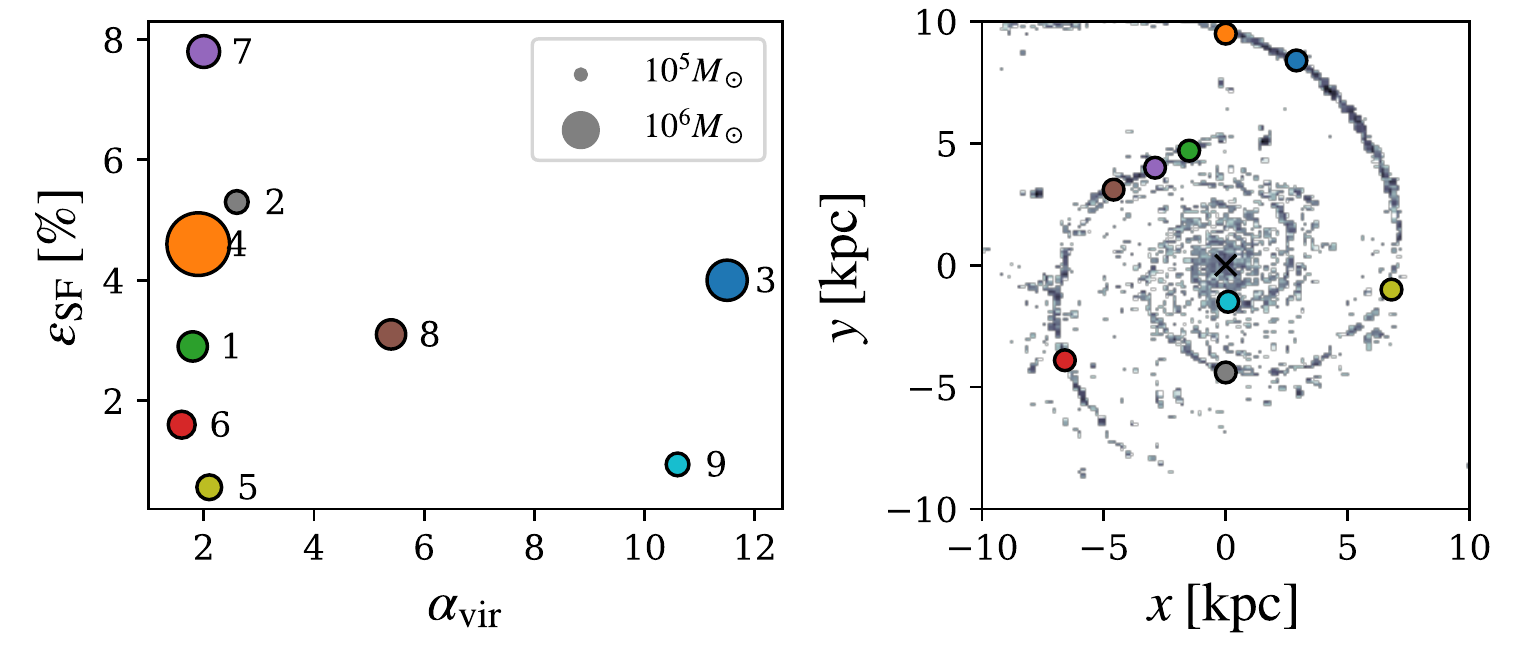}
\caption{Left: the star formation efficiency, $\epsilon_{\rm SF}$ as a function of $\alpha_{\rm vir}$ and mass (size of points) for our 9 example clouds. Right: the locations of these clouds, using the same colours for each cloud as the left panel, over-plotted on the dense gas surface density.}
\label{cloudtrackSF}
\end{centering}
\end{figure}

Star formation efficiencies were estimated for each of our example clouds over a period of 5\,Myr by simply dividing the total mass of stars formed from the component gas particles by the total mass of the cloud. This star formation efficiency is shown in Figure\;\ref{cloudtrackSF} for the tracked clouds. On the left is plotted the star formation efficiency versus the virial parameter measured at the start of the time period used to measure the efficiency. Sizes indicate the cloud masses. In the right panel the locations of each of the GMC's in the disc is shown, with colours consistent between the two panels. Clouds are selected from across the disc, including both arms and a single cloud in the inner disc. There is, surprisingly, no correlation between GMCs in similar regions. For instance, clouds 1, 7 and 8 are in the same arm region but are quite different in terms of $\epsilon_{\rm SF}$ and $\alpha_{\rm vir}$). There is also no correlation with where $\alpha_{\rm vir}$ is highest, with clouds 3 and 4 being very close in the same arm and yet having strikingly different $\alpha_{\rm vir}$. The only trend appears to be with the highest efficiency clouds being located in some of the densest gas in an arm (e.g. Cloud7, 2, 4 and 3), with the medium to low efficiency clouds either in low density regions of an arm (Cloud6 or 5) or off arm entirely (Cloud9). 

These star formation efficiencies range from 1\% to as high as 8\%, but there appears to be no global correlation to $\alpha_{\rm vir}$ nor to $M_c$. However, most of this discrepancy lies with the three highly unbound clouds (Cloud1, Cloud8 and Cloud9), with the remaining clouds with $1<\alpha_{\rm vir}<3$ showing hints of a linear trend of $\epsilon_{\rm SF}$ correlating with $\alpha_{\rm vir}$, but still seemingly consistent with random scatter. As such, it appears that the virial parameter is a poor indicator for the star-forming capacity of these clouds. It may be more suited to describing the individual star forming cores within a cloud, as it can be seen that at $+5$\,Myr there are numerous small clouds that are formed from target clouds after their definition at $\Delta t=0$\,Myr (see Figures \ref{cloudtrack2} to \ref{cloudtrack9}). A more concise assessment of such trends would require a detailed decomposition into different regions which is beyond the scope of this work.

\subsection{Changes in cloud morphology}
\begin{figure}
\includegraphics[trim = 0mm 0mm 0mm 0mm,width=80mm]{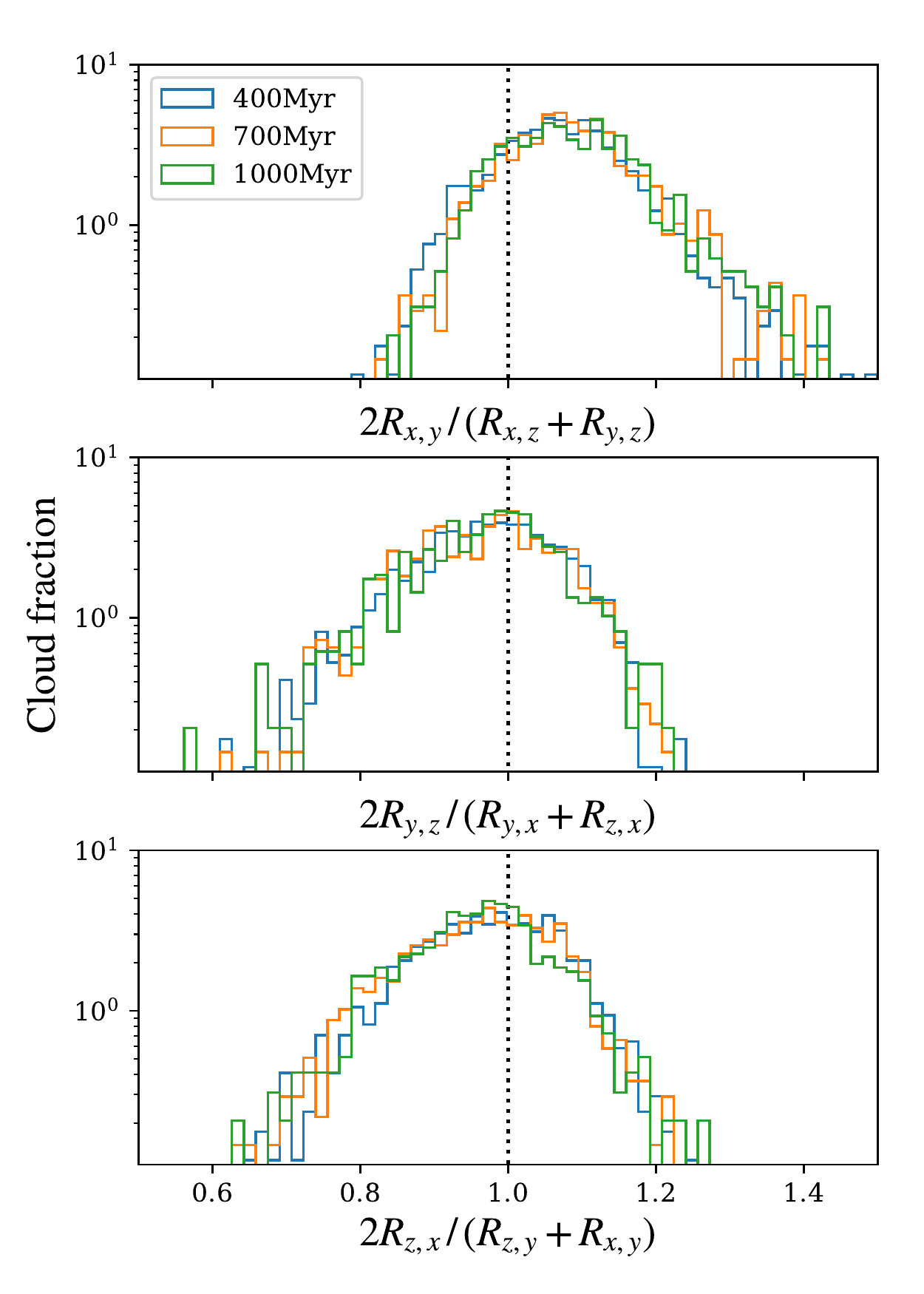}
\caption{Histogram of the fraction of clouds of a given oblateness at three different timeframes in the simulation, measured as the ratio between the measured surface areas in different orthogonal planes. The vertical dashed line indicates clouds that are effectively spherically symmetric.}
\label{gmcface}
\end{figure}

As the tidal force disrupts the galactic disc, clouds are swept into the long bisymmetric spiral pattern. Visual inspection suggests that these GMCs may be different in morphology compared to those in isolation, displaying slight elongations along the strong tidal spiral arms. In Figure\;\ref{gmcface} we show changes in the oblateness of clouds at three different stages in the simulation; just prior to interaction (400\,Myr), at peak spiral response (700\,Myr), and after the companion has long left the system (1000\,Myr). The oblateness is determined by the ratio of the GMC radii calculated in different planes. $R_{x,y}$ is the radius measured when slicing through the $x,y$ plane, measured at the value of $z$ corresponding to the centre-of-mass of the cloud. $2R_{x,y}/(R_{x,z}+R_{y,z})$ therefore defines the asymmetry in the cloud volume with respect to the plane of the galactic rotational angular momentum vector. The vertical dashed line indicates clouds that are effectively spherically symmetric. The data in the upper histogram shows that the clouds are preferentially flattened in the $z$ axis, showing a larger values of $R_{x,y}$ compared to measured in other axes (see the oblateness in other directions in the middle and bottom panels). This general trend persists even in the post-interaction period. However, in the later time-frames (500 and 800\,Myr) there is a very slight shift of clouds to higher values of oblateness in the $x-y$ plane, both in reducing the low and increasing the high value tails. While clouds may be changing morphologically in the presence of the large-scale streaming motions induced by companion passage, the data in Figure\;\ref{gmcface} seem insufficient to fully support this idea (KS-tests between pre and post interaction data give $p$-values of the order of 0.02, significantly higher than the changes in $M_c$ or $\alpha_{\rm vir}$). Previous studies have also seen differences in cloud morphology in the presence of spiral arms, with \citet{2017MNRAS.470.4261D} seeing large filamentary structures just upstream in simulations of density wave driven spiral arms.

A simple calculation of the Jacobi/Roche radii of the GMC's with respect to the companion indicates that even the largest clouds are half the size required to experience tidal destruction effects at 400\,Myr when the companion is closest, and a tenth that at later times (700/1000\,Myr), implying that tidal forces from the companion are not playing a direct role in shaping the clouds.
 
To further investigate the impact of the tidal perturbation on the internal cloud structure, we calculated the z-component of the angular momentum for the clouds as the simulation evolved. Simulations of isolated discs and observations of external galaxies indicate that while most clouds rotate in a prograde manner with disc rotation, there is also a significant population of clouds with retrograde rotation \citep{2003ApJ...599..258R,2008MNRAS.391..844D,2009ApJ...700..358T,2015ApJ...801...33T}. 
We see a fraction of retrograde clouds of around 25\% of the total cloud population when the disc is in isolation, which then rises to 31\% when the spiral is at peak strength, dropping down to pre-interaction values at the end of the simulation. This hints that the clouds are likely experiencing a greater degree of cloud-cloud collisions as a result of a spiral-arm perturbation, which can disrupt clouds into a rotation direction opposed to the net galactic rotation (see Appendix\;\ref{Appx3} for examples of clouds undergoing collisions/mergers). Similar results are seen in \citet{2011MNRAS.417.1318D} who compare cloud rotations with and without an imposed rigidly rotating spiral potential.

\section{Comparison to observations of M51}
\label{SecDisc}

\begin{figure}
\begin{centering}
\includegraphics[trim = 10mm 0mm 0mm 0mm,width=95mm]{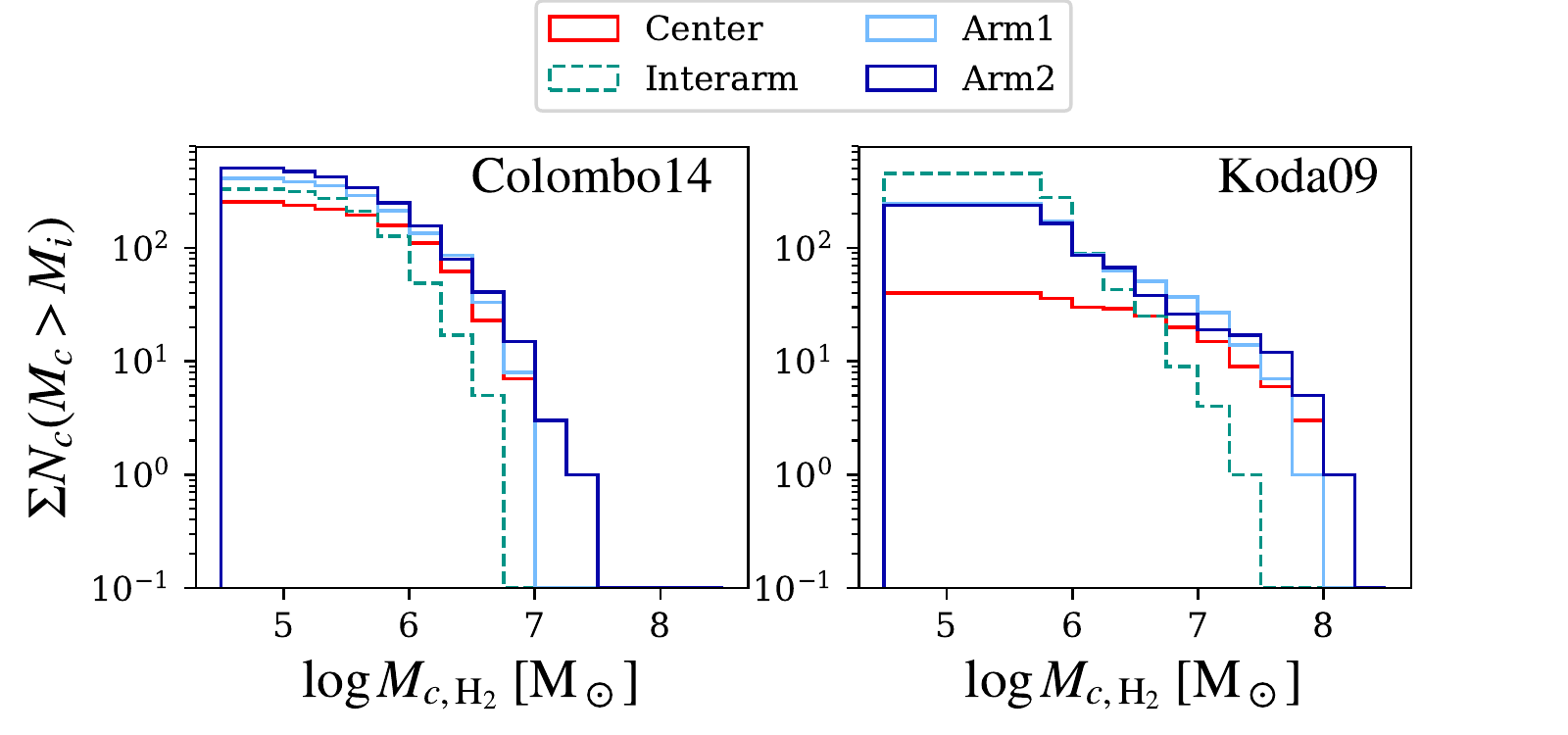}
\caption{Cumulative distribution of M51 GMC masses from \citet{2014ApJ...784....3C} (left) and \citet{2009ApJ...700L.132K} (right). 
Data has been decomposed into the arm definitions of \citet{2017MNRAS.465..460E}.}
\label{PAWSm}
\includegraphics[trim = 10mm 0mm 0mm 0mm,width=95mm]{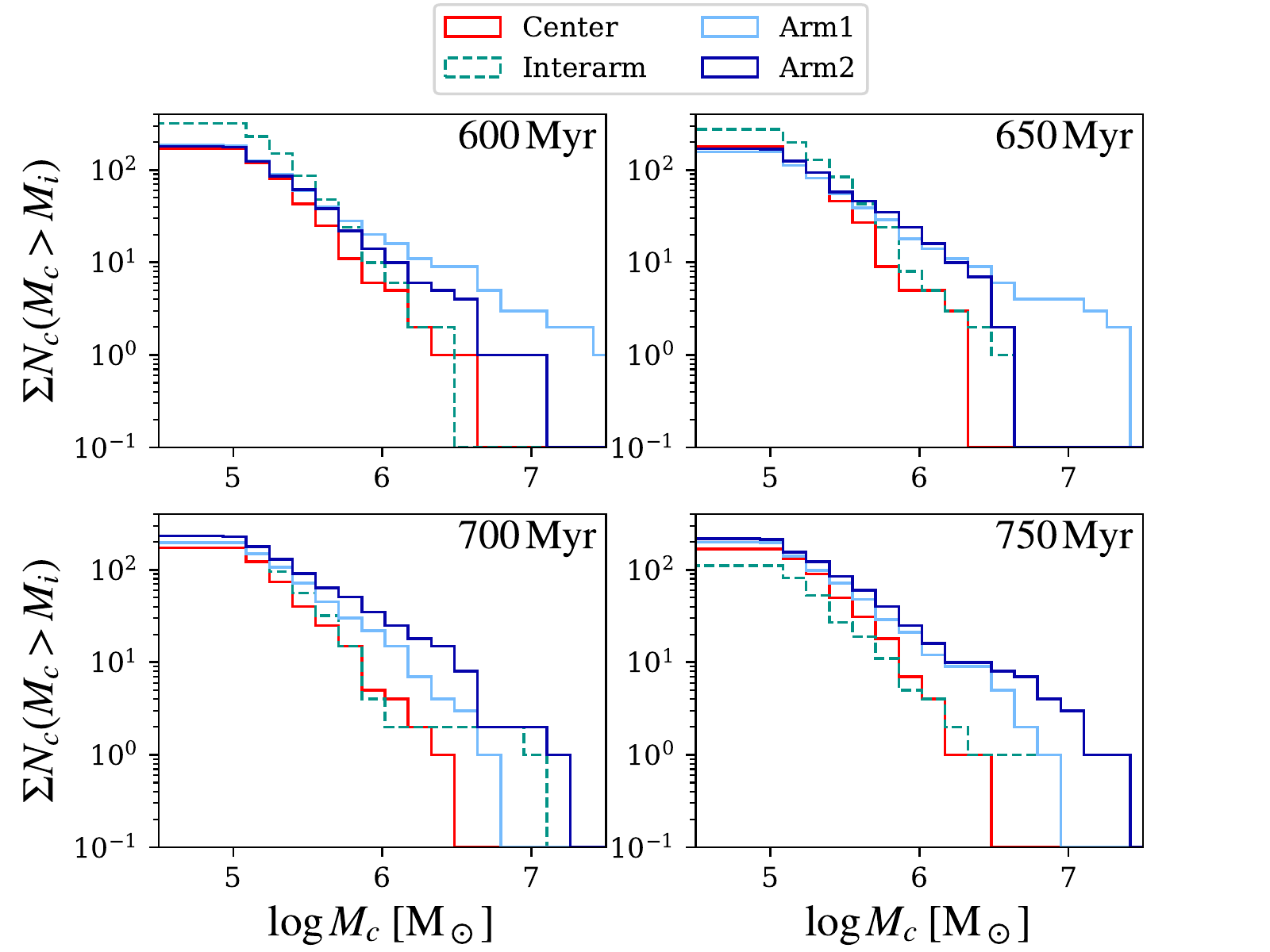}
\caption{Cumulative distribution of GMC masses categorised by region form the tidal spiral simulation presented in this work at four different timeframes. Arm1/2 being analogous to Arm1/2 in Figure\;\ref{PAWSm}, with a similarly defined inter-arm and central region.}
\label{PAWSmineC}
\end{centering}
\end{figure}

We now compare our GMC catalogue with data obtained from observations of M51. We use the PAWS survey catalogue taken from \citet{2014ApJ...784....3C} and the CARMA-derived catalogue of \citet{2009ApJ...700L.132K}. The latter uses data from CO $J=1-0$ survey of the entire M51 disc ($6' \times 8.4'$) to a resolution of 160\,pc to identify a catalogue of 1071 GMCs with a spatial resolution of 160\,pc. The PAWS survey is a 40\,pc resolution survey of CO $J=1-0$ within a narrower field of view focussing only on the inner arm region ($ 4.5' \times 2.8'$), identifying 1508 clouds, offering the largest extragalactic GMC catalogue to date. However, \citet{2014ApJ...784....3C} use a highly segregated categorisation for their GMCs, including many different regions of a single arm. We instead categorise their data according to the definition used in \citet{2017MNRAS.465..460E} where the disc is divided into Arm1 (tail arm), Arm2 (bridge arm, connecting to M51b), inter-arm regions and inner disc. The cumulative histogram of the masses is shown in Figure\;\ref{PAWSm}. Note that clouds from the PAWS catalogue are limited to $R<6$\,kpc, and we have chosen to limit our cloud catalogue to $R<4\,$arcmin in the \citet{2009ApJ...700L.132K} dataset. The latter is because clouds further out are outside of the arm definition in \citet{2017MNRAS.465..460E} and arm structures begin to stray from a log-spiral structure due to their increasing proximity to NGC\,5195. The arms host the largest number of clouds in both datasets, with the inner disc having a similar distribution to the arm/central regions but harbouring fewer mid-mass ($\approx 10^5\,{\rm M_\odot}$) clouds in comparison. Arm2 also appears to harbour higher mass clouds than Arm1 in both datasets.

In Figure\,\ref{PAWSmineC} we show our simulated GMC catalogue at four different times sub-divided into similar categories as Figure\;\ref{PAWSm}. {Four timeframes are shown due to the time-varying nature of both the galaxy and the cloud population}. We use the arm definition of P2017, see their Fig.\,10 and 11 and text for arm definition details. We do not use the exact same definition as \citet{2017MNRAS.465..460E} for two reasons. The first is that the evolving system is quite hard to fit with one underlying model, with arms wrapping up and branching over time. The second is that the disc is not morphologically a perfect match to M51. The spiral arms do not penetrate as far into the centre of the model as the real M51, but do propagate to larger radii, due to the choice of rotation curve (see \citealt{2018MNRAS.474.5645P} for the impact of different rotation curves on tidal spiral properties). We use the same nomenclature and denote Arm2 as the bridge arm that connects to the companion at closest approach.

The GMCs in the simulated disc show some similarities to the observed data. Most notably the inter-arm regions have the lowest fraction of high mass GMCs, consistent with the observed clouds, especially at later times (750\,Myr). The arms tend to have the higher fraction of high mass clouds, with a slight preference for Arm2 hosting the most massive clouds at 700/750\,Myr, which is consistent with the observational data (especially that of  \citealt{2009ApJ...700L.132K}). The simulation does not capture the low mass end of the GMC distribution, which is due to the resolution of the simulation limiting the minimum extractible GMC mass. Azimuthally averaged gas surface densities range from 2\,${\rm M_\odot pc^{-2}}$ to 30\,${\rm M_\odot pc^{-2}}$ from 0 to 10\,kpc, whereas the total gas surface density in the real M51 reaches somewhat higher vales; 6\,${\rm M_\odot pc^{-2}}$ to 70\,${\rm M_\odot pc^{-2}}$ over the same radial range \citep{2007A&A...461..143S}. As such surface densities, and indeed the surface profile itself, have not been tailored to match M51 explicitly it is not surprising that our simulated cloud population does not exactly match that of M51.

One difference between the simulated and observed data is that the simulated clouds do not reach the higher masses seen in M51. A possible explanation for this could simply be the overlap of emission in the observed M51 clouds, with resolution being a known factor in determining cloud distributions (e.g. \citealt{2009ApJ...699L.134P,2016ApJ...831...16L}). However, the mean cloud separation ranges from 100\,pc--150\,pc (with lower values in the perturbed disc compared to the isolated one), larger than the PAWS resolution (40\,pc), so this is unlikely to be the case. We believe it is primarily the lower total gas budget in the simulated galaxy that is responsible for the lower cloud masses compared to observations.

We remind the reader that our simulations do not take into account molecular gas, so direct comparisons to observed clouds properties, which are based off molecular emission, should not be expected to agree exactly (e.g. observed mass profiles should be increased to around $\times 2$; \citealt{2009ApJ...700..358T}).

\section{Conclusions}
\label{SecConc}
We have analysed a simulation of a tidally perturbed disc galaxy with the aim of identifying changes in the GMC population as a result of the interaction. We find that the cloud population undergoes noticeable changes compared to the disc in isolation. The cloud population exhibits a reduction in the small, low mass, well bound clouds after the interaction, continuing to change well after the companion has passed closest approach. This results in the average cloud being are more massive and also less well bound than their isolated counterparts. As such, clouds with $\alpha_{\rm vir}>2$ are not uncommon, as the biggest change the interaction induces is to increase the velocity dispersion of the clouds. The slope of the mass function therefore flattens to favour a high mass cloud population. Small clouds are seemingly agglomerated into large unbound complexes as gas streams into the spiral arms, resulting in a dearth of smaller and well bound clouds that are more common in the pre-interaction case. The interaction also increases the fraction of the galactic gas reservoir contained within the clouds, though the post-interaction phase also sees a gradual drop in the number of clouds. The latter appears to be a result of the reduced time clouds clouds spend within inter-arm regions as the spiral arms wind up, and the gradual inflow of gas/clouds into the galactic centre.

The tidal spiral arms offer safe havens for massive cloud growth by shielding them from the strong shear inherent to the differentially rotating disc. As such, the arms host the high mass and poorly bound clouds almost exclusively, while well-bound clouds can survive the more dangerous inter-arm regions, and are poorer tracers the underlying spiral pattern. By tracking a few example clouds, it is seen that their evolution history is highly diverse. Some are slowly grown from neighbouring gas accumulation, others from collisions with nearby clouds or converging filaments. Clouds can be destroyed abruptly from stellar feedback, or dissipated more slowly as they leave the spiral arms. There appears however to be little correspondence to the star formation efficiency and virial parameter in these example clouds, nor even to their locations such as within the same arm.

Comparisons to the GMC catalogues of the tidally interacting spiral M51 show several similarities, though not an exact match. Both simulations and observations see a preference of more massive clouds in the arms, with observations tending to favour the bridge arm as harbouring the most massive clouds. In our simulation the two arms seem to change in terms of hosting the most massive clouds, though the timestamp at 300\,Myr after perigalacticon passage shows close similarities to the GMCs in M51's arms (and also a similar by-eye morphology). In light of this, a future study is in preparation focused on creating higher resolution simulations specifically tailored to model the M51 system (adapting the M51 model of \citealt{2010MNRAS.403..625D}), where we aim to quantitively compare simulated cloud catalogues to the observed cloud population.

\section*{Acknowledgments}
We utilised the \textsc{pynbody} package \citep{2013ascl.soft05002P} for post-processing and analysing of the \textsc{tipsy} files created by \textsc{gasoline2}. We thank the PAWS team for making their data and catalogues publicly available \citep{2013ApJ...779...42S,2014ApJ...784....3C,2014ApJ...784....4C,2013ApJ...779...43P}, and Jin Koda for the CARMA derived GMC catalogue of M51 \citep{2009ApJ...700L.132K}. We thank K. Wada for useful discussions related to this work. Special thanks to the authors of \textsc{gasoline2} \citep{2017MNRAS.471.2357W} for making this study possible.

\bibliographystyle{mn2e}


\appendix
\section[]{Calculating GMC radii}
\label{Appx1}
We tried numerous approaches for how to define the radius of a GMC boundary. The first, and simplest approach was simply to define the radius as half the maximum distance between the GMC gas particles. While this gave what appeared to be a reasonable value, it was sensitive to the occasional outlying particle. The second was to bound the cloud particles with a convex hull using the \textsc{Qhull} algorithm \citep{2013ascl.soft04016B} or a concave hull using a Delaunary alpha-shape \citep{alphashapes}, in 2D or 3D. However, due to the nature of SPH, the ``elastic band'" approach of such hulls meant that much of the particle mass was outside of the cloud volume, leading to what appeared to be underestimates of the cloud radius. We finally settled on a simpler approach, whereby we re-construct a 3D volume containing only the SPH particles defining the cloud, applying their intrinsic smoothing to neighbouring cells by the smoothing kernel and individual $h$ values. The volume was then collapsed into a 2D column density in each dimension, whereby the surface of the cloud was bound by a contour. We originally adopted $N_{\rm col,GMC}=1\times10^{21}\,{\rm cm^{-2}}$ for a density threshold of $50\,{\rm cm^{-3}}$. However, this was picking up only a few hundred GMCs due to our modest resolution. As such we investigated lower density thresholds of 40$\,{\rm cm^{-3}}$ and 30$\,{\rm cm^{-3}}$ and linearly decreasing $N_{\rm col,GMC}$ accordingly to compensate. We found that 40$\,{\rm cm^{-3}}$ was a good compromise between giving a good number of clouds and not characterising many smaller cloud features into overly large complexes. $N_{\rm col,GMC}$ is effectively another free parameter in the analysis, and was chosen to enclose at least one $h$ for each SPH particle in the cloud. 

Once such areas are defined for the cloud, a radius was computed by considering it analogous to a circle of equivalent area \citep{2009ApJ...699.1092H}. A mean cloud radius was calculated by the average of the three different surface areas, while differences between these values give an indication of the asymmetry of the clouds.

It is worth noting that the results using the contour analysis were remarkably close to what was given by the convex hull value, where it seems the error in bounding by the centre of each SPH particle is balanced by the overestimated area due to the lack of any concave features. The alpha-shape offers what may be the most robust measurement of the intricate structure of the cloud, if not for the missing mass. A volume in keeping with the SPH method could be defined by smoothing out the particle masses into a 3D cube and fitting the alpha-shape to the resulting structure, again defined by some threshold. However this method was proving computationally complex and was deemed too refined an approach for the somewhat corse resolution used in the global simulation.

\section[]{Fits to GMC mass functions}
\label{Appx2}
\begin{figure}
\begin{centering}
\includegraphics[trim =5mm 0mm 0mm 0mm,width=80mm]{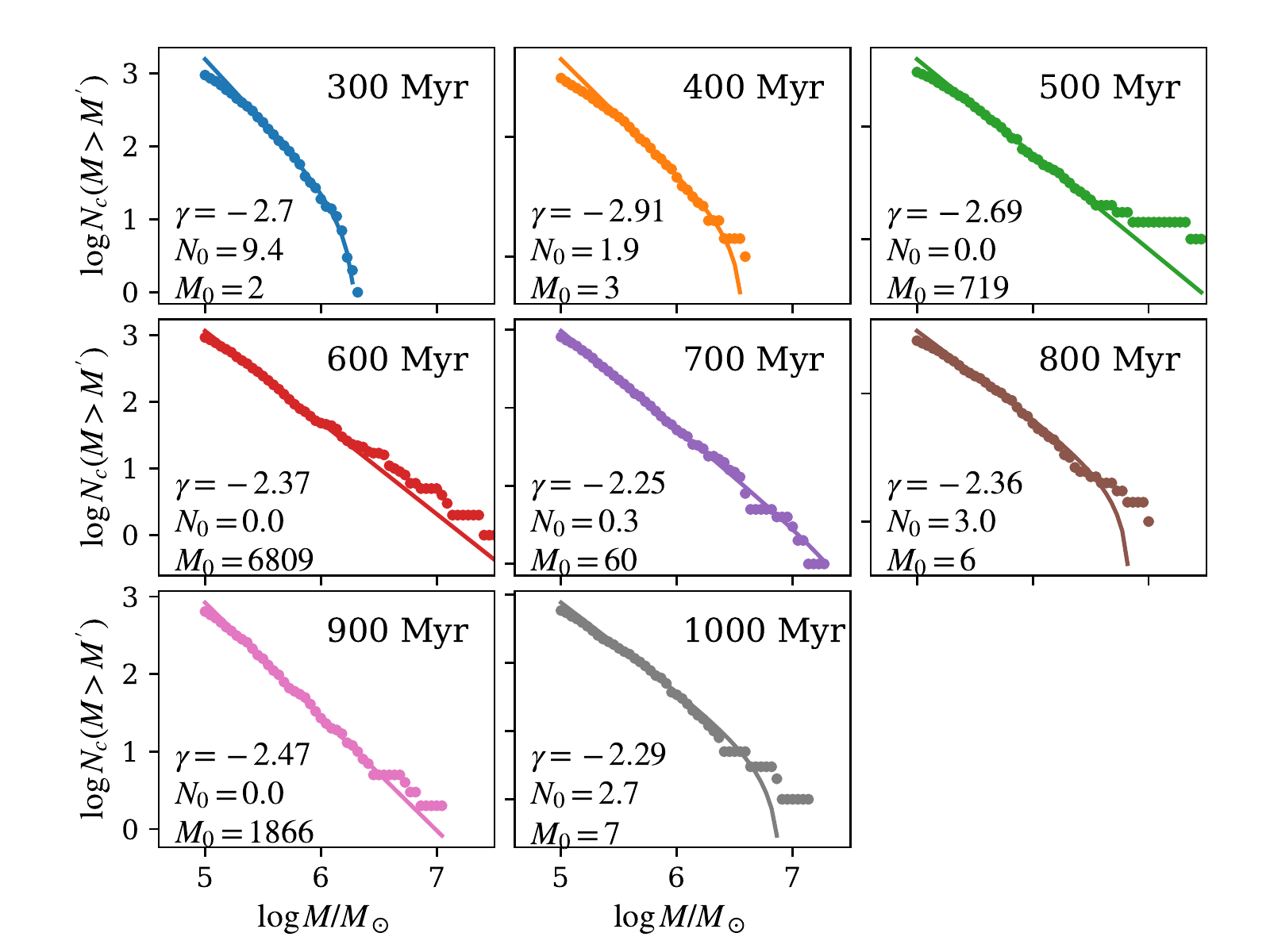}
\caption{Fits to the mass spectra of clouds masses in the model discussed in the main text, where the interaction phase occurs after 400\,Myr. The best-fit parameters are shown in the bottom left of each panel, where $M_0$ is given in units of $10^6\,M_\odot$.}
\label{GMCfit}
\end{centering}
\end{figure}

In Figure\;\ref{GMCfit} we show the fits to the cumulative mass functions defined by Equation\;\ref{massCDF} to the cloud populations in our simulation. Each panel shows a different time-frame. The isolated disc is effectively the 300\,Myr and 400\,Myr timeframes, with the 500--1000\,Myr epoch indicating the tidally perturbed disc. Values for the fitted parameters ($N_0$, $M_0$ and the slope of the mass function, $\gamma$) are shown in the bottom-left of each panel.

Interestingly some of the slopes show a closer resemblance with a pure power law rather than a truncated one (with a low $N_0$ combined with a high value of $M_0$, beyond the end of our high-mass tail), especially in the times after the interaction. This is seen in some galactic environments, such as the inter-am regions of M51 \citep{2014ApJ...784....3C}, M33, and the outer disc of the Milky Way \citep{2005PASP..117.1403R}. It is thus puzzling why M51, which closer resembles the interaction model than the Milky Way or M33, shows clearer truncations. We leave an in-depth modelling of the M51 system specifically to a future work, where we will investigate the changing shape of the mass spectra in different regions to see if a similar environmental dependance is seen as in \citet{2014ApJ...784....3C}, where arm regions show clear cut-offs compared to inter-arm regions.

\section[]{Tracking select individual clouds}
\label{Appx3}
Here we show an additional 8 GMCs tracked in our interacting disc, each displaying a different evolutionary history.

\begin{figure*}
\begin{centering}
\includegraphics[trim = 10mm 10mm 20mm 0mm,width=180mm]{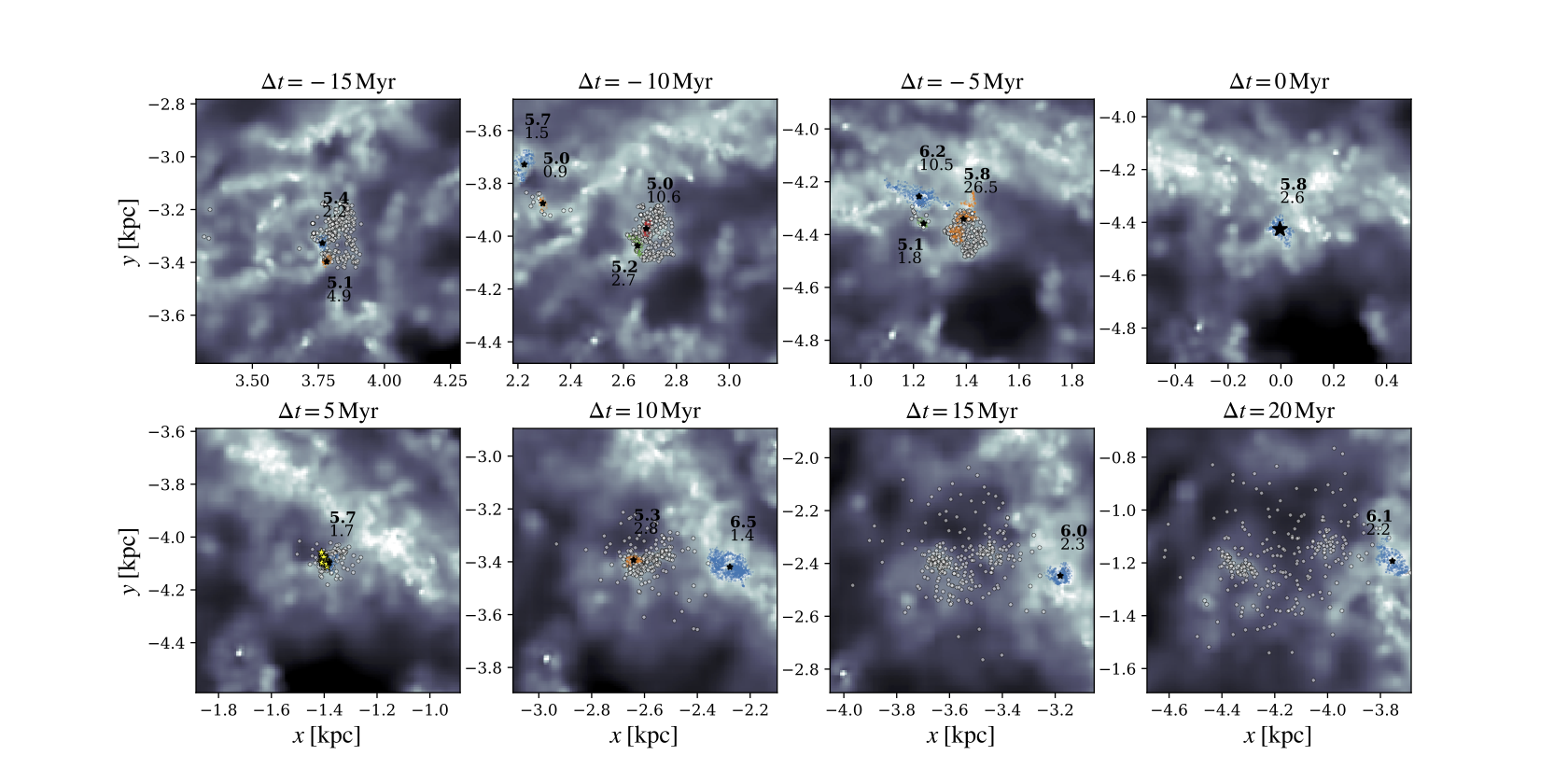}
\caption{As Figure \ref{cloudtrack1} but for Cloud2. Cloud2 is defined as it leaves the spiral arm, dissipating as it converts mass into a cluster of star partilces.}
\label{cloudtrack2}
\end{centering}
\end{figure*}

\begin{figure*}
\begin{centering}
\includegraphics[trim = 10mm 10mm 20mm 0mm,width=180mm]{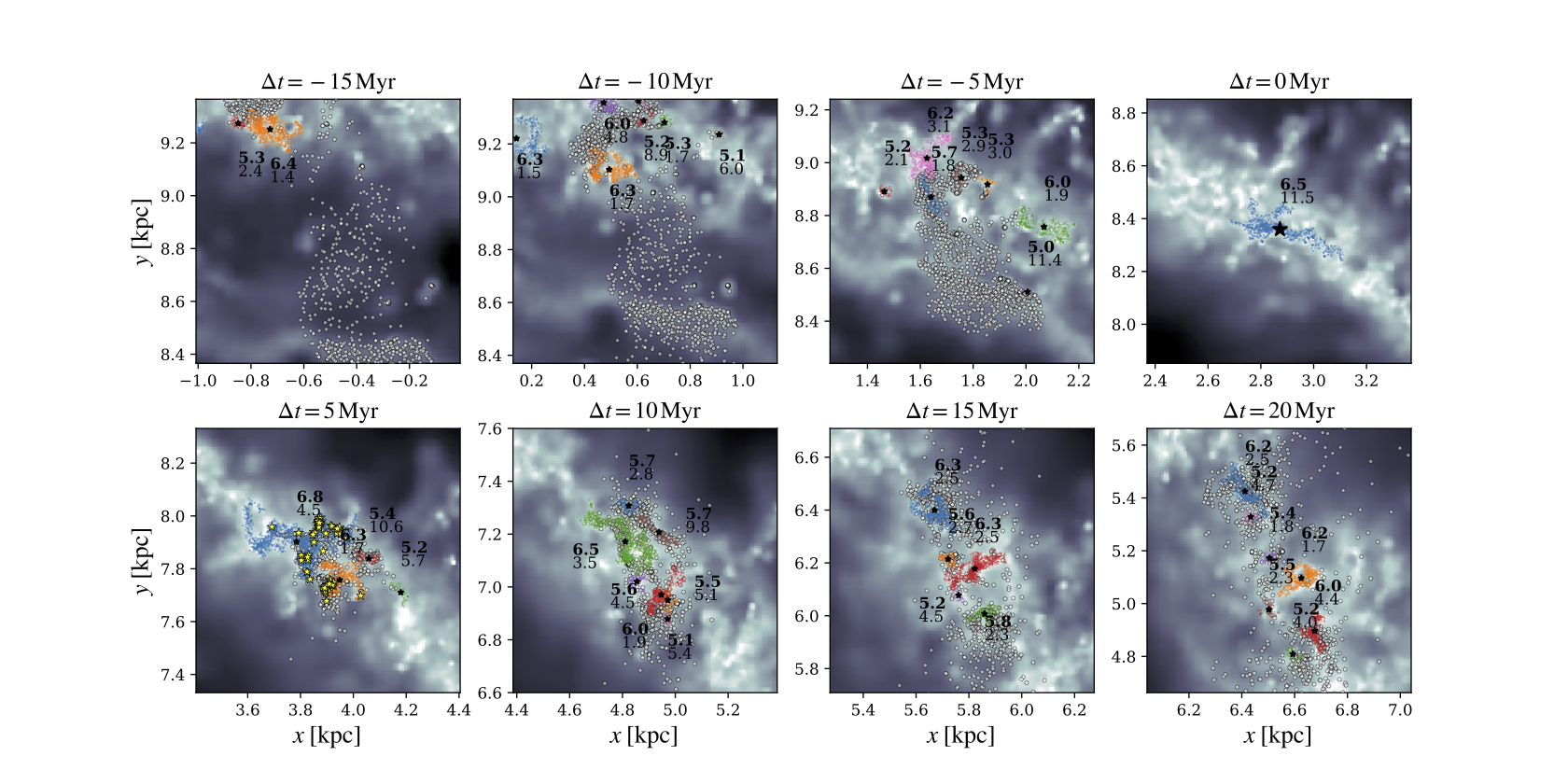}
\caption{As Figure \ref{cloudtrack1} but for Cloud3. Cloud3 is mid-way on a primary arm, and is triggered by the incident ridge of gas upstream.}
\label{cloudtrack3}
\end{centering}
\end{figure*}

\begin{figure*}
\begin{centering}
\includegraphics[trim = 10mm 10mm 20mm 0mm,width=180mm]{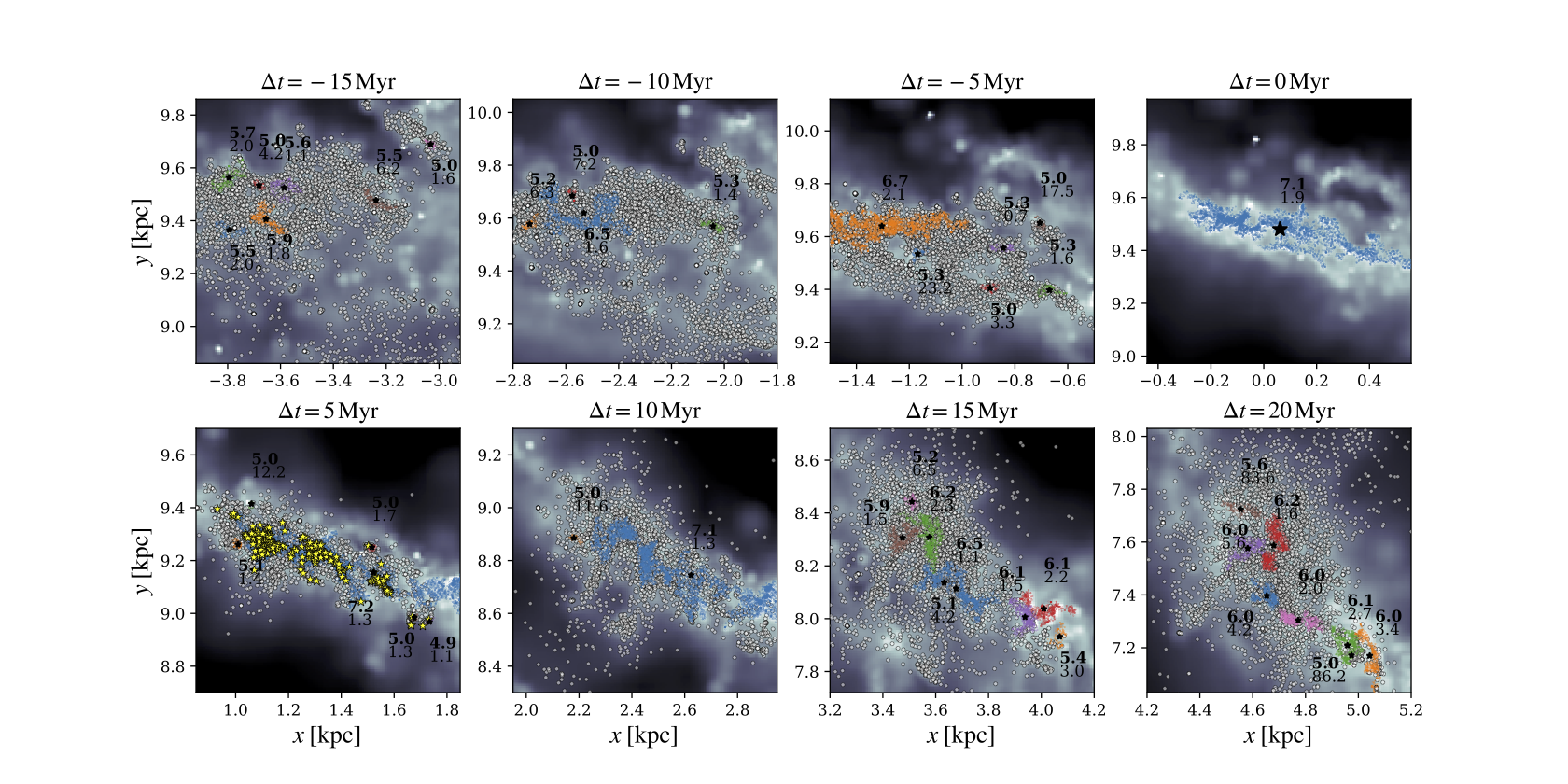}
\caption{As Figure \ref{cloudtrack1} but for Cloud4. Cloud4 is near the end of a primary arm. It is one of the largest clouds in our sample, and appears to be made up of gas from a wide range of environments.}
\label{cloudtrack4}
\end{centering}
\end{figure*}

\begin{figure*}
\begin{centering}
\includegraphics[trim = 10mm 10mm 20mm 0mm,width=180mm]{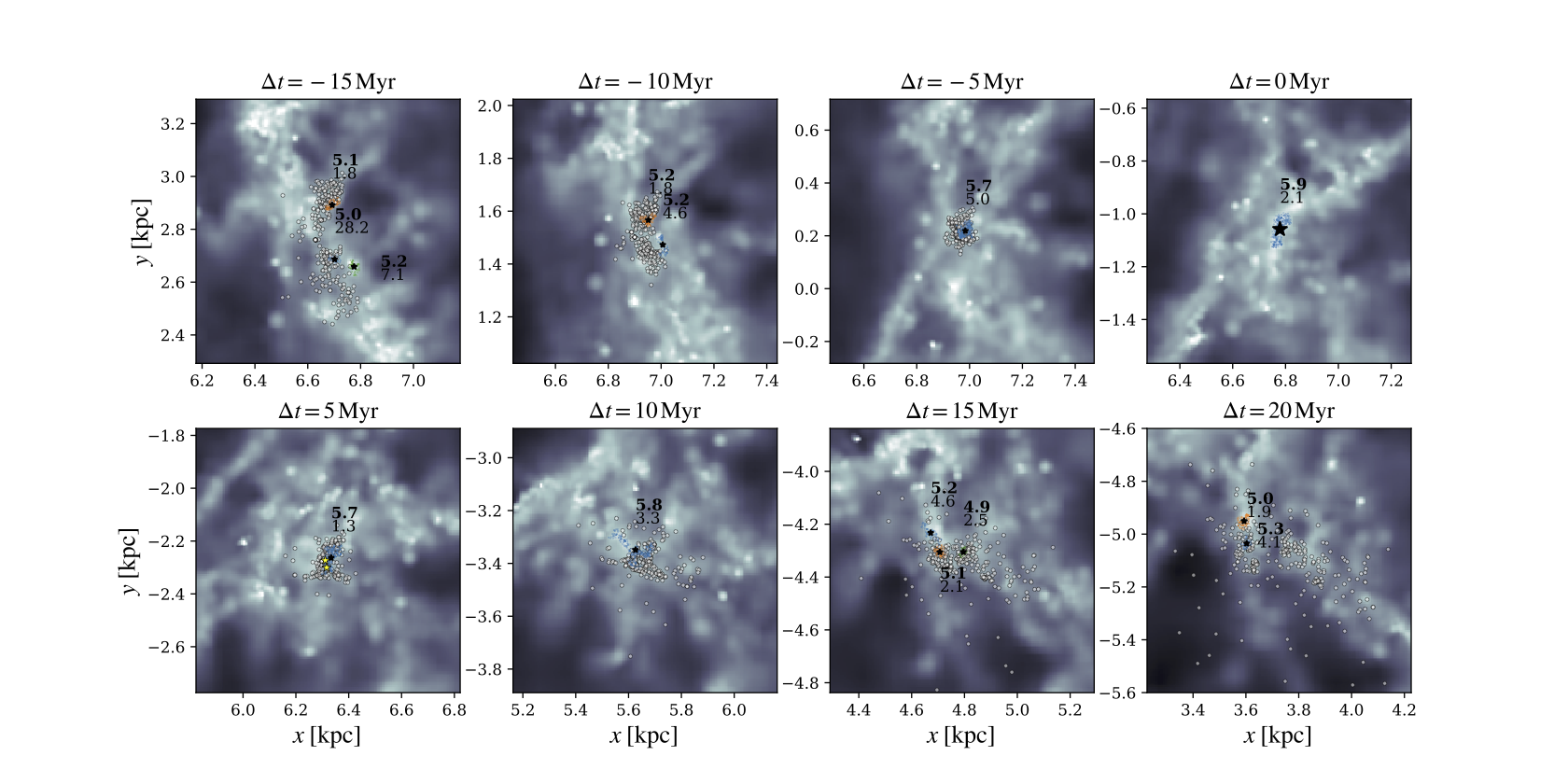}
\caption{
As Figure \ref{cloudtrack1} but for Cloud5. Cloud5 is a massive but nearly-bound cloud that shears apart into a spur and smaller inheritor clouds as it leaves the spiral arm.}
\label{cloudtrack5}
\end{centering}
\end{figure*}

\begin{figure*}
\begin{centering}
\includegraphics[trim = 10mm 10mm 20mm 0mm,width=180mm]{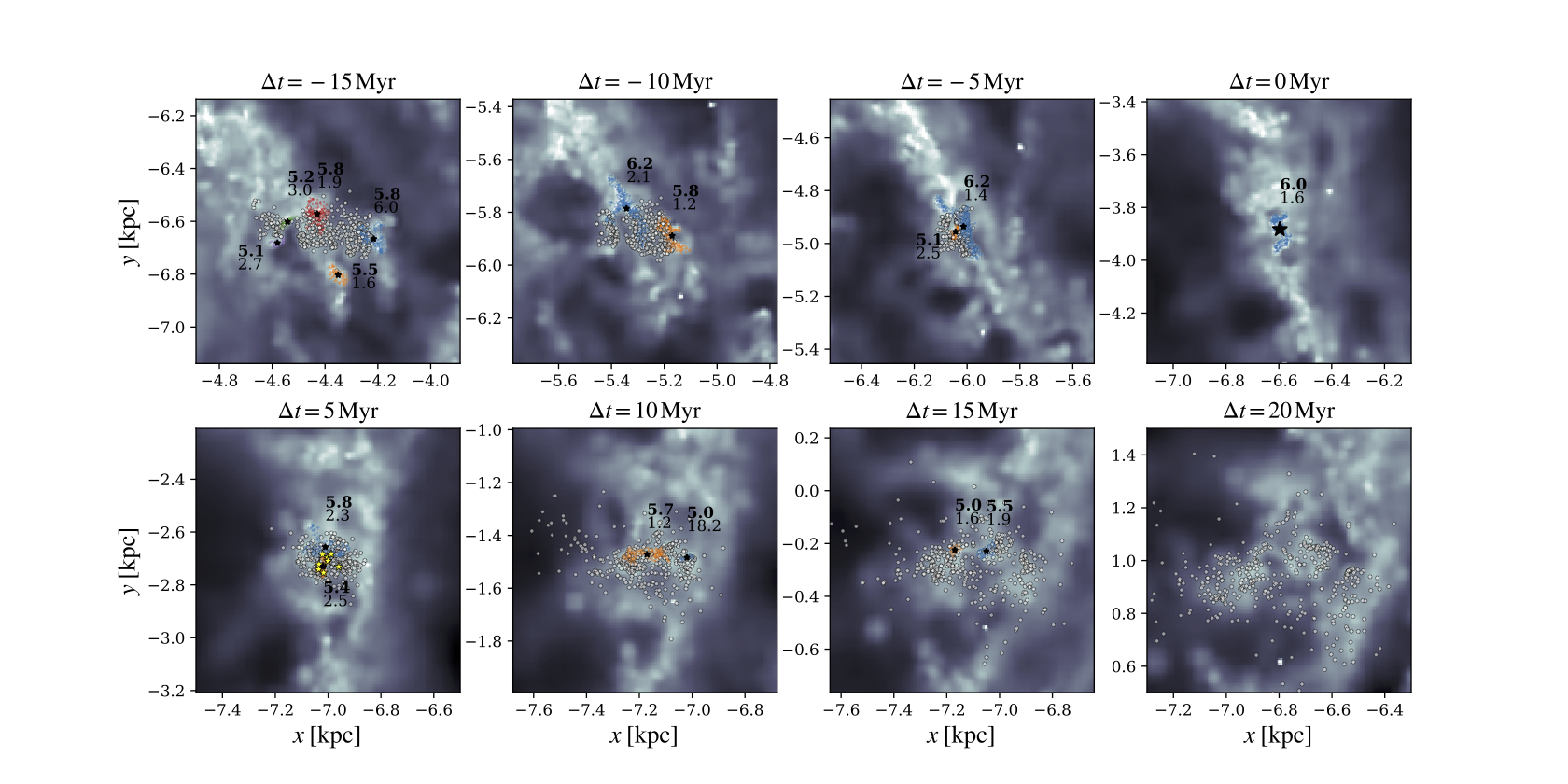}
\caption{As Figure \ref{cloudtrack1} but for Cloud6. Cloud6 lies the end of an arm and is made of two minor clouds that are slowly sheared apart.}
\label{cloudtrack6}
\end{centering}
\end{figure*}

\begin{figure*}
\begin{centering}
\includegraphics[trim = 10mm 10mm 20mm 0mm,width=180mm]{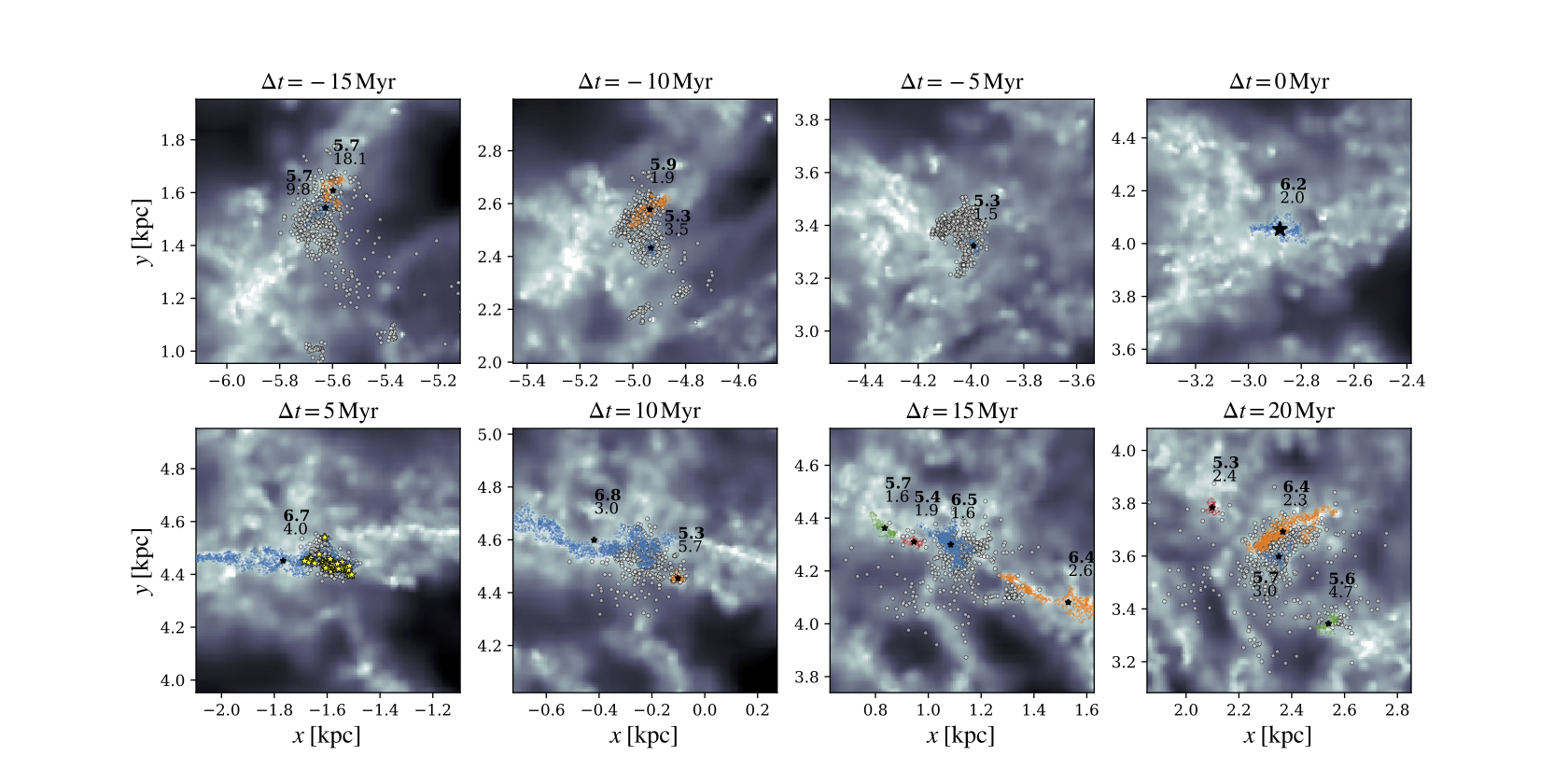}
\caption{As Figure \ref{cloudtrack1} but for Cloud7. Cloud7 is moves into the spiral pattern near the centre of the disc, experiencing a strong shock and large burst in star formation.}
\label{cloudtrack7}
\end{centering}
\end{figure*}

\begin{figure*}
\begin{centering}
\includegraphics[trim = 10mm 10mm 20mm 0mm,width=180mm]{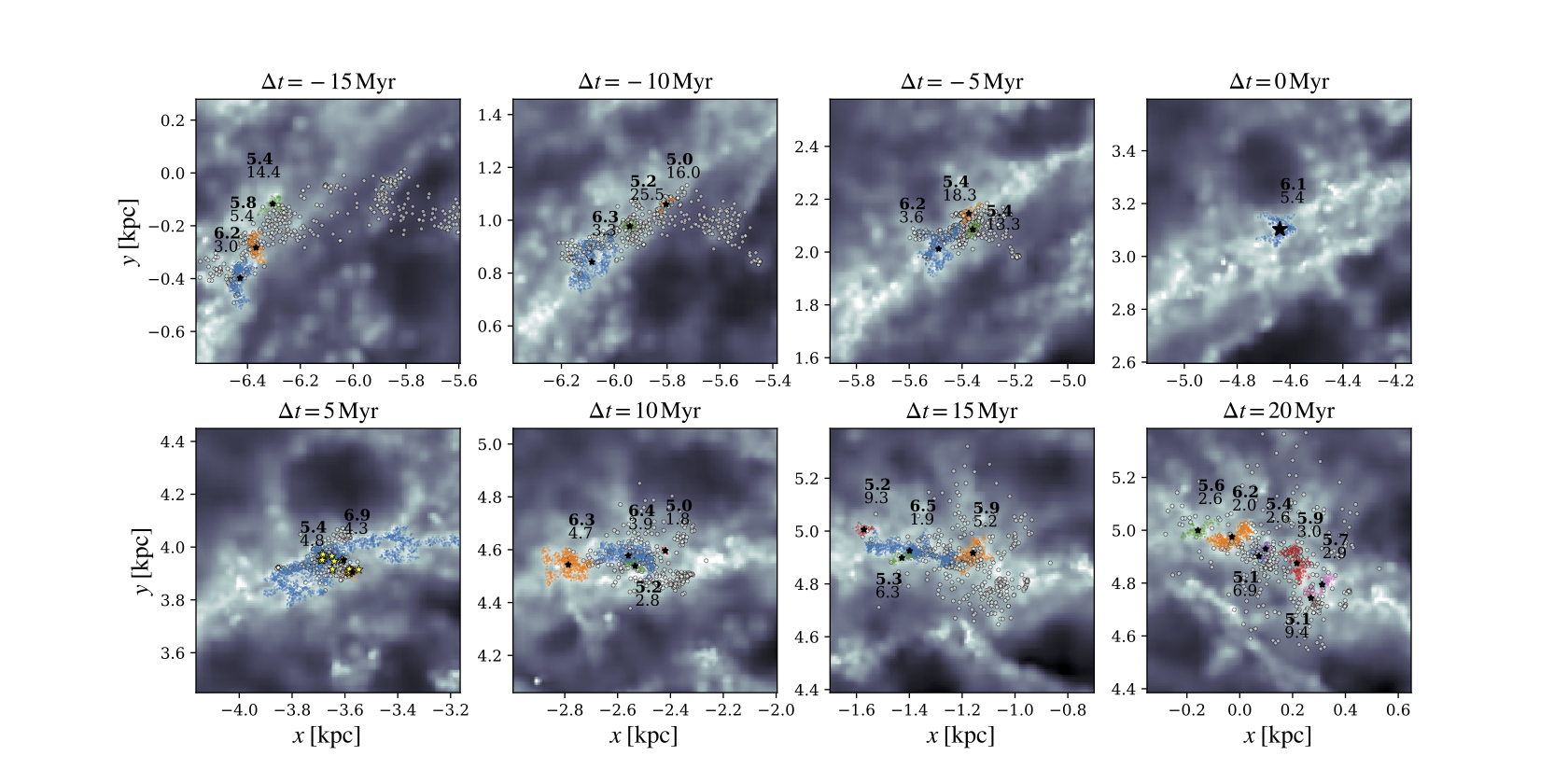}
\caption{As Figure \ref{cloudtrack1} but for Cloud8. Cloud8 is a coalescence of many smaller clouds and gas streaming into the arm, that fragments into many smaller clouds while still within the arm.}
\label{cloudtrack8}
\end{centering}
\end{figure*}

\begin{figure*}
\begin{centering}
\includegraphics[trim = 10mm 10mm 20mm 0mm,width=180mm]{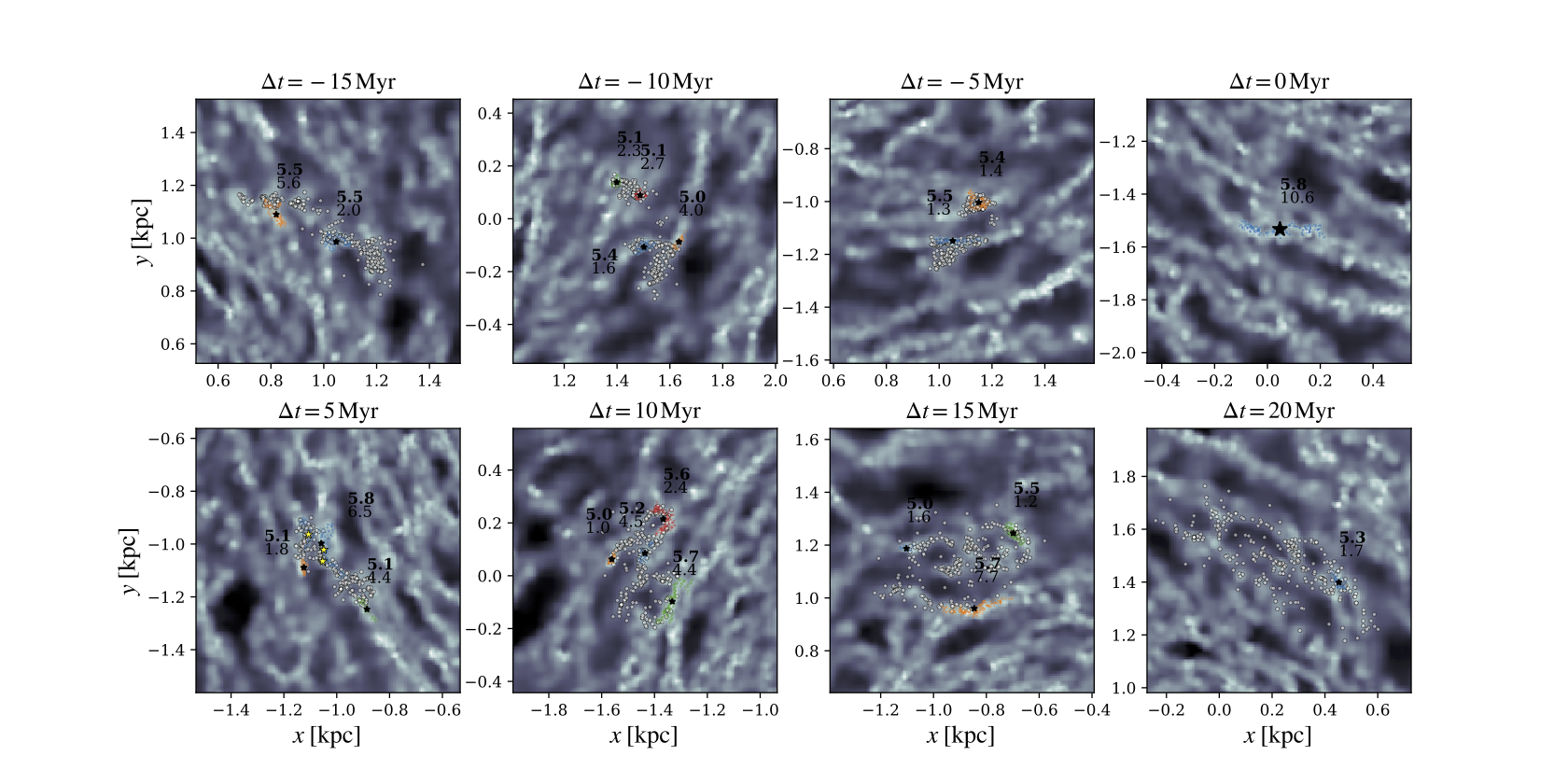}
\caption{As Figure \ref{cloudtrack1} but for Cloud9, a cloud in the inner region of the disc that is highly unvirialised and, unlike the other clouds, far away from the 2-armed spiral potential.}
\label{cloudtrack9}
\end{centering}
\end{figure*}

Cloud2 (Fig. \ref{cloudtrack2}) appears to be just leaving an arm at the time of definition, having been formed from material that made up a number of clouds some 15\,Myr in the past. As it leaves, it forms a cluster of star particles, having the second highest star formation efficiency out of our example clouds. Free of the spiral arm, and having lost  a significant portion of its mass, the cloud dissipates into the diffuse inter-arm ISM.

The evolution of Cloud3 (\ref{cloudtrack3}) shows the interesting case of a clear triggering of structure by the collision with an incident ridge of gas. This ridge is not defined as a cloud, and is instead a long thin filament that shocks a number of smaller clouds in the arm into a single object. This shocked ensemble of gas immediately forms a number of stars, which then breaks apart into a number of inheritor clouds.

In Figure\;\ref{cloudtrack4} we show one of the largest clouds in our sample: Cloud4. This cloud is made of a great swath of gas, both from low density regions streaming into the arms and constituents of other small clouds. The cloud is surprisingly well bound considering its prior dispersive state, both in terms of $\alpha_{\rm vir}$ and how it maintains its structure at later times, only truly breaking apart 20\,Myr later. The cloud, or rather filament, experiences a moderate level of star formation along its length (an efficiency of nearly 5\%), aiding to break apart the cloud into 7 individual clouds later on before it leaves the arm.

Cloud5 (Fig.\;\ref{cloudtrack5}) is a small compact cloud, with a relatively low virial parameter. It has a quieter evolution than the clouds already shown, and dissipates as it leaves the spiral arm. Gas is lost as it moves into the inter-arm region as an short spur, with a portion of the gas remaining in a couple of smaller clouds.

Cloud6 lies on the end of the tail arm that is being gradually wound up (Fig.\;\ref{cloudtrack6}) . Here a number of smaller clouds appear to form two moderate sized objects that interact with each other for nearly 30\,Myr before breaking apart. The two clouds at -10\,Myr could be prime examples of an off-centre collision, with the S-shaped cloud at 0\,Myr showing the clouds rotating around a common centre of mass and making up our target cloud. The star formation efficiency in this cloud is the lowest of those shown, at a level of only 1.5\% despite being quite massive. Eventually the mutual tidal forces on each of the clouds, coupled with shearing motion as they leave the arm, result in their dissipation.

In contrast, Cloud7 experiences the strongest levels of relative star formation of our clouds (Fig.\;\ref{cloudtrack7}), reaching an efficiency of 7.8\%. The early stages of the cloud show it begins on an inter-arm spur, before being coincident with the spiral at 0\,Myr. There seems no clear reason why this cloud produces so many stars, other than it seems to have accreted a large amount of gas that converges on its position over the previous 10\,Myr, including cannibalising the orange cloud seen at $\Delta t=-10$\,Myr. The feedback from the substantial amount of young stars blows gas out into the surrounding media, though the approach of a long filament-type GMC from $\Delta t =5$\,Myr onwards acts to repopulate the region with gas, resulting in clouds composed of the same gas as Cloud7 to survive the arm passage.

Cloud8, shown in Figure\;\ref{cloudtrack8}, shows three clouds combining within a spiral arm, being fuelled by the upstream gas entering from upstream. The resulting cloud is quite poorly bound, and quickly becomes part of a long filament. The interesting aspect of this cloud is that only 20\,Myr later the gas once contained within now belongs to a number of clearly defined smaller clouds, aligned in a ``beads on a string" kind of manner. Most of these resultant clouds are very weakly bound, and will not survive the transition to the inter-arm region.

The final cloud, Cloud9, is very different to those already shown, as it resides within the inner disc as opposed to the spiral arms. The cloud is very loosely bound, and splits its gas budget into several inheritor clouds over a short timescale.


\bsp
\label{lastpage}
\end{document}